\documentclass[10pt,floatfix,%
twocolumn,
amsmath,amssymb,
aps,
superscriptaddress,
prapplied,
]{revtex4-2}

\usepackage[svgnames,dvipsnames]{xcolor}
\usepackage{url}
\usepackage{graphicx}
\graphicspath{{.}{figs/}{../figs/}}
\usepackage{afterpage}
\usepackage{dcolumn}
\usepackage{bm}
\usepackage{tikz}
\usetikzlibrary{quantikz2,positioning,intersections,angles,quotes}
\usepackage{adjustbox}
\usepackage[caption=false]{subfig}
\usepackage{comment}
\usepackage{physics}
\usepackage{mathtools}

\usepackage[colorlinks,citecolor=DarkGreen,linkcolor=FireBrick,linktocpage,unicode,hypertexnames=false]{hyperref}
\usepackage[capitalize]{cleveref}
\crefname{enumi}{}{}
\crefname{section}{Sec.}{Sec.}
\usepackage{autonum}

\begin{document}
\title{Data-Efficient Error Mitigation for Physical and Algorithmic Errors \\ in a Hamiltonian Simulation
}
\author{Shigeo Hakkaku}
\email{shigeo.hakkaku@ntt.com}
\affiliation{NTT Computer and Data Science Laboratories, NTT, Inc., 3-9-11 Midori-cho, Musashino, Tokyo 180-8585, Japan}
\affiliation{NTT Research Center for Theoretical Quantum Physics, NTT, Inc., 3-1 Morinosato-Wakamiya, Atsugi 243-0198, Japan}

\author{Yasunari Suzuki}
\affiliation{NTT Computer and Data Science Laboratories, NTT, Inc., 3-9-11 Midori-cho, Musashino, Tokyo 180-8585, Japan}
\affiliation{NTT Research Center for Theoretical Quantum Physics, NTT, Inc., 3-1 Morinosato-Wakamiya, Atsugi 243-0198, Japan}

\author{Yuuki Tokunaga}
\affiliation{NTT Computer and Data Science Laboratories, NTT, Inc., 3-9-11 Midori-cho, Musashino, Tokyo 180-8585, Japan}
\affiliation{NTT Research Center for Theoretical Quantum Physics, NTT, Inc., 3-1 Morinosato-Wakamiya, Atsugi 243-0198, Japan}

\author{Suguru Endo}
\email{suguru.endou@ntt.com}
\affiliation{NTT Computer and Data Science Laboratories, NTT, Inc., 3-9-11 Midori-cho, Musashino, Tokyo 180-8585, Japan}
\affiliation{NTT Research Center for Theoretical Quantum Physics, NTT, Inc., 3-1 Morinosato-Wakamiya, Atsugi 243-0198, Japan}
\affiliation{JST, PRESTO, 4-1-8 Honcho, Kawaguchi, Saitama, 332-0012, Japan}

\begin{abstract}
Quantum dynamics simulation via Hamiltonian simulation algorithms is one of the most crucial applications in the quantum computing field.
While this task has been relatively considered the target in the fault-tolerance era, the experiment for demonstrating utility by an IBM team simulates the dynamics of an Ising-type quantum system with the Trotter-based Hamiltonian simulation algorithm with the help of quantum error mitigation \cite{kimEvidenceUtilityQuantum2023}.
In this study, we propose the data-efficient 1D extrapolation method to mitigate not only physical errors but also algorithmic errors of Trotterized quantum circuits in both the near-term and early fault-tolerant eras.
Our proposed extrapolation method uses expectation values obtained by Trotterized circuits, where the Trotter number is selected to minimize both physical and algorithmic errors according to the circuit's physical error rate.
We also propose a method that combines the data-efficient 1D extrapolation with purification QEM methods, which improves accuracy more at the expense of multiple copies of quantum states or the depth of the quantum circuit.
Using the 1D transverse-field Ising model, we numerically demonstrate our proposed methods and confirm that our proposed extrapolation method suppresses both statistical and systematic errors more than the previous extrapolation method.
\end{abstract}
\maketitle
\section{Introduction}
One of the most promising applications of a fault-tolerant quantum computer is simulating dynamics of many-body problems, such as condensed matter physics and quantum chemistry~\cite{childsFirstQuantumSimulation2018,yoshiokaHuntingQuantumclassicalCrossover2024}.
Hamiltonian simulation algorithms have been mainly discussed in the regime of fault tolerance, and numerous improvements have been made in this area~\cite{lloydUniversalQuantumSimulators1996,suzukiGeneralizedTrotterFormula1976,campbellRandomCompilerFast2019,berryExponentialImprovementPrecision2014,berrySimulatingHamiltonianDynamics2015,babbushExponentiallyMorePrecise2016,lowHamiltonianSimulationQubitization2019}.
Although most such algorithms are beyond the capacity of the current quantum computers, the Trotter-based Hamiltonian simulation algorithm has already been performed in the current quantum device by an IBM team, with the utility of quantum devices being investigated~\cite{kimEvidenceUtilityQuantum2023}.

In the above experiment, the error-extrapolation quantum error mitigation (QEM)\cite{temmeErrorMitigationShortDepth2017,liEfficientVariationalQuantum2017,endoPracticalQuantumError2018} has played a crucial role in improving the accuracy of the simulation result by suppressing the physical errors due to coupling to the environment.
Besides the near-term application, it has been pointed out that QEM can significantly improve the computation accuracy of quantum algorithms in the early fault-tolerant quantum computing (FTQC) era and contribute to reducing resources such as code distances and T gate counts~\cite{suzukiQuantumErrorMitigation2022,xiongSamplingOverheadAnalysis2020,piveteauErrorMitigationUniversal2021,lostaglioErrorMitigationQuantumAssisted2021,wahlZeroNoiseExtrapolation2023}.

While the primary target of QEM has been the suppression of physical errors, the computation accuracy of quantum algorithms is also restricted by algorithmic errors because of the insufficiency of the circuit depth. Ref.~\cite{endoMitigatingAlgorithmicErrors2019} elaborated a QEM method tailored to 
noisy Trotterized quantum circuits to suppress both the algorithmic and physical errors.
This method uses error-extrapolation QEM methods to mitigate physical error rates and then applies the algorithmic error extrapolation method for Trotter-based quantum simulation by leveraging the outcomes for fewer Trotter steps.
Although this method does not impose any additional hardware requirements and may enhance computation accuracy even in current quantum devices, the variance of its estimator is relatively large.
This is because the method performs two successive extrapolations, thereby requiring many noisy expectation values.

In this work, we propose a data-efficient error extrapolation method for physical and algorithmic errors, which is an even more robust QEM for Trotter-based quantum algorithms.
The proposed extrapolation method uses expectation values obtained by Trotterized circuits where the Trotter number is selected to minimize both physical and algorithmic errors as a function of the physical error rate of the circuit. 
Our proposed method performs only one extrapolation according to the physical noise,
and it gives a more accurate result than sequentially applying extrapolation for physical and algorithmic errors.
To identify such a Trotter number, we clarify the dependence of the optimal Trotter numbers on the error rates of physical noise by evaluating the distance between the exact time-evolved operator and the noisy Trotterized circuit.
Then, by analyzing an output state of the noisy Trotterized circuit with such a Trotter number, we discuss the coefficients of the proposed extrapolation method.

We also propose a method that combines the proposed data-efficient error extrapolation with purification QEM methods \cite{koczorExponentialErrorSuppression2021,koczorDominantEigenvectorNoisy2021,hugginsVirtualDistillationQuantum2021,caiResourceefficientPurificationbasedQuantum2021}, which improves accuracy even more at the expense of the number of copies of quantum states or the depth of the quantum circuit.
We dub this method as Trotter subspace expansion.
The Trotter subspace expansion ensures that the error-mitigated computation outcome leads to the physical one.
We construct the error-mitigated state $\rho_\text{QEM}=\frac{\rho_\text{extra}^2}{\text{Tr}[\rho_\text{extra}^2]}$ for the effective state of the proposed extrapolation method $\rho_\text{extra}=\sum_{i} c_{i} \rho(p_i) $ ($c_{i} \in \mathbb{R}$) by the help of the purification-based QEM methods.
Here, $\rho(p_i)$ are noisy output states with $p_i$ being the different physical error rates, where the Trotter number is chosen to minimize the physical and algorithmic errors.

To benchmark our proposed methods, we numerically compare them with the existing error mitigation method proposed in Ref.~\cite{endoMitigatingAlgorithmicErrors2019} under finite measurements by considering the dynamics of 1D transverse field Ising model with 10 qubits.
We find that the MSE of the data-efficient extrapolation is the least in some existing QEM methods if $10^8\leq N \leq 10^{11}$.
Also, the numerical result shows that the Trotter subspace expansion would be more useful than the data-efficient extrapolation if we could use at least $10^{12}$ number of measurements.
Its converged MSE is 710 times smaller than the raw data and 31 times smaller than the data-efficient extrapolation.
Thus, if we can execute many quantum computers in parallel, the Trotter subspace expansion would be more useful than the other QEM methods.

\section{Preliminaries}\label{sec:preliminaries}
In this section, we review the error extrapolation method for Trotter errors \cite{endoMitigatingAlgorithmicErrors2019}.
Then, we explain purification-based quantum error mitigation methods, which use multiple copies of noisy quantum states.
We combine these methods with our proposed extrapolation method, as we will see later in \cref{subsec:our_proposed_extrapolation}.

\subsection{Trotterization}\label{sec:trotterization}
Here, we give a brief explanation of the Trotterization \cite{suzukiGeneralizedTrotterFormula1976,lloydUniversalQuantumSimulators1996}.
The Trotterization approximates an exponential of a sum of operators, such as a time evolution operator, by a product of elementary exponentials.
Thus, it is used as a subroutine of many quantum algorithms, including Gibbs state preparation \cite{poulinSamplingThermalQuantum2009}, Hamiltonian simulation \cite{lloydUniversalQuantumSimulators1996}, and phase estimation \cite{nielsenQuantumComputationQuantum2010,linHeisenbergLimitedGroundStateEnergy2022}.
In the following, we explain the Trotterization using a time evolution operator as an example.
Let $H=\sum_jH_j$, where each $H_j$ acts on a constant number of qubits.
The time evolution operator $e^{-iHt}$ can be approximated by a product of unitaries $e^{-it/MH_j}$:
\begin{align}
    \exp(-iHt) = \bqty{\prod_j \exp\pqty{-i\frac{t}{M}H_j}}^M + \mathcal{O}\pqty{\frac{t^2}{M}}, \label{eq:Trotter}
\end{align}
where $M$ is called the Trotter number or Trotter step and $\mathcal{O}\pqty{\frac{t^2}{M}}$ is an algorithmic error.
Therefore, $M$ determines the level of algorithmic error. If quantum circuits are error-free, the algorithmic error can be arbitrarily suppressed as the Trotter number increases. Denoting $\epsilon_M = 1/M$ and $U_{\epsilon_M}$ as
\begin{align}
    U_{\epsilon_M}\pqty{t}\coloneqq \bqty{\prod_j \exp\pqty{-i\epsilon_M t H_j}}^{1/\epsilon_M},
\end{align}
we can obtain
\begin{align}
    \lim_{\epsilon_M\to +0} U_{\epsilon_M}\pqty{t} = \exp\pqty{-iHt} \label{eq:continuity_of_Trotter}.
\end{align}

However, we cannot increase $M$ infinitely because current quantum devices suffer from noise. In such a device, a number of Trotter steps introduces more physical errors, and thus, there exists the optimal Trotter number $M_\text{opt}$ \cite{kneeOptimalTrotterizationUniversal2015}. Denoting the ideal and noisy quantum process of one layer of the Trotterization, i.e., $\prod_j \exp\pqty{-i\frac{t}{M}H_j} $ as $\mathcal{U}_\text{TR}$ and $\mathcal{E}_\text{TR}$, respectively, the distance between the target process $\mathcal{U}_\text{targ} \coloneqq \bqty{e^{-iHt}}$, where $\bqty{A}(\sigma) = A \sigma A^\dag$, and the noisy Trotterized process reads:

\begin{align}
D\pqty{ \mathcal{U}_\text{targ} ,  \mathcal{E}_\text{TR}^M } &\leq M D(\mathcal{U}_\text{targ}^{\frac{1}{M}}, \mathcal{E}_\text{TR}) \\
&\leq M \Bqty{ D\pqty{\mathcal{U}_\text{targ}^{\frac{1}{M}}, \mathcal{U}_\text{TR}} + D\pqty{\mathcal{E}_\text{TR},\mathcal{U}_\text{TR}} }.
\label{Eq: distoftwoprocesses}
\end{align}
Here,
$D\pqty{\mathcal{E}_1, \mathcal{E}_2} = \text{max}_\sigma \norm{\mathcal{E}_1 \pqty{\sigma} -   \mathcal{E}_2 \pqty{\sigma}_1}$
is the distance between two quantum processes $\mathcal{E}_1$ and $\mathcal{E}_2$ for the trace norm $\norm{A} =\text{Tr}\bqty{\sqrt{A^\dag A}}$ with $\sigma$ being the properly chosen quantum states \cite{gilchristDistanceMeasuresCompare2005}.
Note that we assume the noise process for each layer is the same.
In the first line of \cref{Eq: distoftwoprocesses}, we use the chaining property $D\pqty{\mathcal{E}_1 \circ \mathcal{E}'_1, \mathcal{E}_2 \circ \mathcal{E}'_2} \leq D\pqty{\mathcal{E}_1, \mathcal{E}'_1}+D\pqty{\mathcal{E}_2, \mathcal{E}'_2}$.
In the second line, we use the triangle inequality.
Since we have $D(\mathcal{U}_\text{targ}^{\frac{1}{M}}, \mathcal{U}_\text{TR}) \leq \frac{a}{M^2}$ and $D\pqty{\mathcal{E}_\text{TR},\mathcal{U}_\text{TR} } \leq b$ for some constants $a, b >0$, we get:
\begin{align}
    D\pqty{ \mathcal{U}_\text{targ} ,  \mathcal{E}_\text{TR}^M } \leq \frac{a}{M} + b M, 
\end{align}
which indicates that there is an optimized Trotter number $M_\text{opt}=\sqrt{\frac{a}{b}}$.
See Refs.~\cite{kneeOptimalTrotterizationUniversal2015,endoMitigatingAlgorithmicErrors2019} for details.

\subsection{Trotter Error Extrapolation}
In this section, we review how to suppress the algorithmic error introduced in \cref{sec:trotterization} according to Ref.~\cite{endoMitigatingAlgorithmicErrors2019}.
This method suppresses the algorithmic error by regarding an inverse of the Trotter number as an ``error rate'' in the standard polynomial error extrapolation method.

Suppose that one applies the unitary operator $U_{\epsilon_M}\pqty{t}$ to the state $\ket{\psi}$, then the expectation value of an observable $A$ is given by
\begin{align}
    \ev{A\pqty{t}}\pqty{\epsilon_M} = \ev{U^\dag_{\epsilon_M}\pqty{t}A U_{\epsilon_M}\pqty{t}}{\psi}.
\end{align}
Expanding $\ev{A\pqty{t}}\pqty{\epsilon_M}$ as a function of $\epsilon_M$ gives
\begin{align}
    \ev{A\pqty{t}}\pqty{\epsilon_M} = \ev{A\pqty{t}}\pqty{0} + \sum_{i=1}^{n'}A\pqty{t}_i \epsilon_M^i + \mathcal{O}\pqty{\epsilon_M^{n'+1}} \label{eq:expansion_as_a_function_of_Trotter_steps},
\end{align}
where $\ev{A\pqty{t}}\pqty{0} = \ev{\exp\pqty{iHt}A\exp\pqty{-iHt}}{\psi}$ i.e., the algorithmic-error-free expectation value of $A$, and $\ev{A\pqty{t}_i}$ stand for the real coefficients of the expansion.

From \cref{eq:continuity_of_Trotter,eq:expansion_as_a_function_of_Trotter_steps}, we can confirm that the error extrapolation method can be applied with algorithmic errors by reducing the Trotter number.
Normally, we increase the Trotter step up to $M_\text{opt}$.
Combining the set of data points with different Trotter steps $\Bqty{\ev{A\pqty{t}}\pqty{\epsilon_{M_i}}}_{i=0}^{n'}$, where $M_{n'}\leq \cdots M_1 \leq M_0 = M_\text{opt}$, with \cref{eq:expansion_as_a_function_of_Trotter_steps}, we obtain
\begin{align}
    \ev{A}_\text{est}\pqty{0}
    &= \sum_{i=0}^{n'}\ev{A}\pqty{\epsilon_{M_i}}\prod_{k\neq i}\frac{\epsilon_{M_k}}{\epsilon_{M_k}-\epsilon_{M_i}}\\
    &= \ev{A}\pqty{0} + \mathcal{O}\pqty{\epsilon^{n'+1}_{M_\text{opt}}}.
\end{align}
Thus, we see that the algorithmic error can be suppressed to $\mathcal{O}\pqty{\epsilon^{n'+1}_{M_\text{opt}}}=\mathcal{O}\pqty{M_{n'}^{-\pqty{n'+1}}}$.

The variance of the estimator $\ev{A}_\text{est}\pqty{0}$ is given by
\begin{align}
    \text{Var}\bqty{A_\text{est}\pqty{0}} = \sum_{i=0}^{n'}\text{Var}\bqty{A\pqty{\epsilon_{M_i}}}\pqty{\prod_{k\neq i}\frac{\epsilon_{M_k}}{\epsilon_{M_k}-\epsilon_{M_i}}}^2.
\end{align}
Then, the required sampling cost of the estimator within the additive error $\delta$ is given by $\mathcal{O}\pqty{\delta^{-2}\text{Var}\bqty{A_\text{est}\pqty{0}}}$.

\section{Data-efficient physical and algorithmic extrapolation}\label{subsec:our_proposed_extrapolation}
Here, we introduce a data-efficient physical and algorithmic error extrapolation method, which suppresses statistical errors more than the previous method \cite{endoMitigatingAlgorithmicErrors2019}.
In this method, we select data points on a line in the plane with axes representing physical error and inverse of Trotter error.
In the following, we will explain the relationship between the optimal Trotter number $M_\text{opt}$ and the error rate of the global depolarizing noise $p_\text{global}$, i.e., $M_\text{opt} \propto 1/\sqrt{p_\text{global}}$, thereby providing a  1D extrapolation method.

\subsection{Optimal Trotter Number}
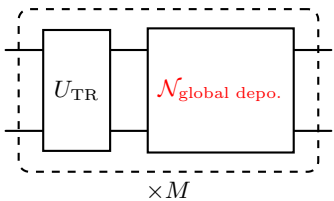
\begin{figure}[tb]
\centering
    \begin{quantikz}
        &\gate[2]{U_\text{TR}}\gategroup[2, steps=2, style={dashed, rounded corners, inner xsep=6pt}, label style={label position=below, anchor=north, yshift=-0.2cm}]{$\times M$} & \gate[2, label style=red]{\mathcal{N}_\text{global depo.}} &\\
        & & &
    \end{quantikz}
    \caption{
        Schematic circuit for deducing an optimal Trotter number as a function of the error rate of the global depolarizing noise.
        $U_\text{TR}$ is one layer of the Trotterization, $M$ is the Trotter number, and $\mathcal{N}_\text{global depo.}$ is the global depolarizing noise.
    }
    \label{fig:schematic_circuit_for_deducing_opt_Trotter_num}
\end{figure}
Suppose that the global depolarizing channel $\mathcal{N}_\text{global depo.}$ occurs after one layer of the Trotterization of the $n$-qubit target process $\mathcal{U}_\text{TR}$ as shown in \cref{fig:schematic_circuit_for_deducing_opt_Trotter_num}.
The global depolarizing noise is defined as
\begin{align}
    \mathcal{N}_\text{global depo.} \pqty{\rho} = \pqty{1-p_\text{global}} \rho + \frac{p_\text{global}}{2^n} I^{\otimes n}.
\end{align}
Although we use the global depolarizing channel for simplicity, recent studies \cite{qinErrorStatisticsScalability2023,tsubouchiUniversalCostBound2023} show that the local noise in a quantum circuit can be approximated as the global depolarizing noise in the large-depth regime theoretically and numerically.
Therefore, the relationship that we will see later is also expected to be valid in the large-depth regime.

From the chaining property, the second term on the right-hand side of \cref{Eq: distoftwoprocesses} follows that 
\begin{align}
    D\pqty{\mathcal{E}_\text{TR},\mathcal{U}_\text{TR}}
    &= D\pqty{\mathcal{N}_\text{global depo.} \circ\mathcal{U}_\text{TR},\mathcal{U}_\text{TR}} \\
    & =  D\pqty{\bqty{I^{\otimes n}}, \mathcal{N}_\text{global depo.}}\label{eq:distance_between_one_layer_noisy_and_ideal_Trotter}
\end{align}
Then, $D\pqty{\bqty{I^{\otimes n}}, \mathcal{N}_\text{global depo.}}$ can be rewritten as
\begin{align}
    &D\pqty{\bqty{I^{\otimes n}}, \mathcal{N}_\text{global depo.}}\\
    =& \max_\sigma\norm{\sigma - \Bqty{\pqty{1-p_\text{global}}\sigma + \frac{p_\text{global}}{2^n}I^{\otimes n}}}\\
    =& p_\text{global}\max_\sigma\norm{\sigma - \frac{1}{2^n}I^{\otimes n}}\label{eq:distance_between_identity_and_depolarizing}.
\end{align}
Substituting \cref{eq:distance_between_identity_and_depolarizing} into \cref{eq:distance_between_one_layer_noisy_and_ideal_Trotter}, we obtain
\begin{align}
    D\pqty{\mathcal{E}_\text{TR},\mathcal{U}_\text{TR}} = p_\text{global} \max_\sigma \norm{\sigma - \frac{1}{2^n}I^{\otimes n}}.
\end{align}
From the above equation, we see that the distance between $\mathcal{E}_\text{TR}$ and $\mathcal{U}_\text{TR}$ could be proportional to $p_\text{global}$, i.e., $D\pqty{\mathcal{E}_\text{TR},\mathcal{U}_\text{TR}} \propto p_\text{global}$.
Thus, an optimized Trotter number could be given by
\begin{align}
    M_\text{opt} = \text{floor}\pqty{\frac{c}{\sqrt{p_\text{global}}}} \label{eq:opt_Trotter_num_depends_on_physical_error_rate}
\end{align}
for a positive constant $c$, which would depend on the number of qubits $n$, the simulation time $t$, and the parameters of a given Hamiltonian, as we will see later.
We expect that if we choose only data points satisfying \cref{eq:opt_Trotter_num_depends_on_physical_error_rate} for a fixed constant $c$, the corresponding extrapolation would suppress both physical and algorithmic errors and also be more data-efficient than the previously proposed two-dimensional extrapolation.

The error rate of the global depolarizing noise could be related to that of the local stochastic noise.
Suppose that a quantum circuit with $N$ quantum gates.
Assuming that the local stochastic noise $\mathcal{N}_\text{local}$ that acts on a quantum state with probability $p_\text{local}$ occurs after each quantum gate.
Then, the probability that a quantum state is not affected by all of the local noise channels per depth would be approximated by $1 - Np_\text{local}$.
Thus, given that all of the local noise channels in a quantum circuit are modeled by one global depolarizing noise, the error rate of global depolarizing noise $p_\text{global}$ is $\mathcal{O}\pqty{Np_\text{local}}$.

By analyzing the Trotter error shown in \cref{eq:remainder} or a tighter bound \cite{childsTheoryTrotterError2021}, we can reduce a more fine-grained optimal Trotter number.
For example, let us consider the dynamics of a 1D transverse-field Ising model
\begin{align}
H_\text{1D TFI} = J\sum_{i=1}^n Z_i Z_{i+1} + B\sum_{i=1}^n X_i,
\end{align}
where we set $Z_{n+1}=Z_1$.
Using the tighter 1st Trotter error with commutator scaling \cite{childsTheoryTrotterError2021}, we obtain
\begin{align}
    \norm{U_\text{targ}^{\frac{1}{M}} - U_\text{TR}}_\text{op}\leq 2\abs{J} \abs{B} n \frac{t^2}{M^2},
\end{align}
where $\norm{A}_\text{op}$ is the spectral norm of $A$.
From this inequality, we expect that $D\pqty{\mathcal{U}_\text{targ},\mathcal{U}_\text{TR}}\leq \tilde{a} \abs{J} \abs{B} n \frac{t^2}{M^2}$ for a constant $\tilde{a}$.
Combining this inequality with $D\pqty{\mathcal{E}_\text{TR}, \mathcal{U}_\text{TR}}\propto p_\text{global}$, we get
\begin{align}
    M_\text{opt} = \order{\abs{t}\sqrt{\frac{\abs{J}\abs{B}n}{p_\text{global}}}}.
\end{align}
If we assume $p_\text{global} = \order{np_\text{local}}$, we obtain
\begin{align}
    M_\text{opt} = \text{floor}\pqty{f \abs{t}\sqrt{\frac{\abs{J}\abs{B}}{p_\text{local}}}}, \label{eq:opt_trotter_1d_tfi}
\end{align}
where $f$ is a positive constant.
In \cref{sec:scaling_of_coeff_of_opt_Trotter_num}, we discuss the validity of the above equation by numerically analyzing the scaling of the optimal Trotter number as a function of the simulation time.

We can also calculate the scaling of the optimal Trotter steps in the case of the 2D TFI Hamiltonian:
\begin{align}
    H_\text{2D TFI} = J \sum_{i\neq j}Z_iZ_j + B \sum_iX_i,
\end{align}
where we consider the periodic conditions.
If we consider the $n \times n$ square lattice, the 1st Trotter error with commutator scaling of the 2D TFI can be obtained as
\begin{align}
    \norm{U_\text{targ}-U_\text{TR}}_\text{op}\leq 4\abs{J}\abs{B}n^2\frac{t^2}{M^2}.
\end{align}
Thus, the optimal Trotter step can be given by
\begin{align}
    M_\text{opt} = \order{\abs{t}\sqrt{\frac{\abs{J}\abs{B}}{p_\text{global}}}}.
\end{align}

\subsection{Extrapolation Method using Optimal Trotter Number}
Now, we provide a data-efficient extrapolation method for both errors using \cref{eq:opt_Trotter_num_depends_on_physical_error_rate}, which we dub data-efficient 1D extrapolation.
Let $\rho_0$ be the initial state of the quantum circuit shown in \cref{fig:schematic_circuit_for_deducing_opt_Trotter_num}.
The output state of the quantum circuit can be given by
\begin{align}
    \rho\pqty{M, p_\text{global}} =& \pqty{1-p_\text{global}}^M \bqty{U_\text{TR}^M}\pqty{\rho_0}\\
    &+ \Bqty{1-\pqty{1-p_\text{global}}^M}\frac{I^{\otimes n}}{2^n}.\label{eq:output_state}
\end{align}
According to Ref.~\cite{lloydUniversalQuantumSimulators1996,sandersQuantumAlgorithmsHamiltonian2007}, the first-order Trotterization can approximate a target time evolution unitary as
\begin{align}
    U_\text{targ} = U_\text{TR}^M + \frac{t^2}{2M}\sum_{i<j} \bqty{H_i,H_j} + \sum_{m=3}^\infty E\pqty{m},
\end{align}
where $E\pqty{m}$ can be upper-bounded by
\begin{align}
    \norm{E\pqty{m}}_\text{op}\leq M\norm{\frac{Ht}{M}}_\text{op}^m \frac{1}{m!} \label{eq:remainder}.
\end{align}
Then, we can rewrite the Trotterization as
\begin{align}
    U_\text{TR}^M = U_\text{targ} - \frac{t^2}{2M}h_\text{comm} - \sum_{m=3}^\infty E\pqty{m},
\end{align}
where $h_\text{comm} \coloneqq \sum_{i<j}\bqty{H_i,H_j}$.
Substituting the above equation into \cref{eq:output_state}, we get
\begin{align}
    \rho\pqty{M,p_\text{global}} =& \pqty{1-p_\text{global}}^M \pqty{\rho_\text{exact} - \frac{t^2}{2M} \Delta + \frac{t^4}{4M^2}\eta - \kappa} \\
    &+ \Bqty{1-\pqty{1-p_\text{global}}^M}\frac{I^{\otimes n}}{2^n} + \mathcal{O}\pqty{\frac{1}{M^3}},
\end{align}
where
\begin{gather}
    \rho_\text{exact} \coloneqq \bqty{U_\text{targ}}\pqty{\rho_0},\quad \Delta \coloneqq h_\text{comm}\rho_0 U_\text{targ}^\dag + \text{h.c.}\\
    \eta \coloneqq \bqty{h_\text{comm}}\pqty{\rho_0}, \quad \kappa\coloneqq \sum_{m=3}^\infty E\pqty{m} \rho_0 U^\dag_\text{targ} + \text{h.c.}
\end{gather}
Here $\norm{\kappa}_\text{op} = \mathcal{O}\pqty{p_\text{global}}$.
Using the Taylor expansion, $\rho\pqty{M,p_\text{global}}$ for small $p_\text{global}$ can be approximated by the following expression:
\begin{align}
    \rho\pqty{M,p_\text{global}} \approx& \pqty{1-Mp_\text{global}} \pqty{\rho_\text{exact} - \frac{t^2}{2M} \Delta + \frac{t^4}{4M^2}\eta - \kappa} \\
    &+Mp_\text{global}\frac{I^{\otimes n}}{2^n} + \mathcal{O}\pqty{\frac{1}{M^3}}.
\end{align}
Substituting \cref{eq:opt_Trotter_num_depends_on_physical_error_rate} into the above expression and considering \cref{eq:remainder}, we have
\begin{align}
    \rho\pqty{p_\text{global}} =& \rho_\text{exact} +  p_\text{global}^{\frac{1}{2}}\pqty{c\frac{I^{\otimes n}}{2^n}-c\rho_\text{exact} - \frac{t^2}{2c}\Delta}\\
    &+ p_\text{global}\pqty{\frac{t^2}{2}\Delta + \frac{t^4}{4c^2}\eta - \frac{\kappa}{p_\text{global}}}\\
    &+ \mathcal{O}\pqty{p_\text{global}^{\frac{3}{2}}}.\label{eq:expansion_of_Trottrized_state}
\end{align}
Here, $p_\text{global}^{-1}\norm{\kappa}_\text{op} = \mathcal{O}\pqty{1}$.
Thus, from \cref{eq:expansion_of_Trottrized_state}, $\rho\pqty{p_\text{global}}$ is expected to be expanded as a Taylor series in $p_\text{global}^{\frac{1}{2}}$, and we expect that the expectation value of an observable $A$ for $\rho\pqty{p_\text{global}}$ is given by
\begin{align}
    \ev{A}\pqty{p_\text{global}} = \ev{A}\pqty{0} + \sum_{i=1}^m A_i p_\text{global}^{\frac{i}{2}} + \mathcal{O}\pqty{p_\text{global}^{\frac{m+1}{2}}} \label{eq:expansion_of_ev_under_global_depo},
\end{align}
where $A_i$ are the real coefficients.
As in a standard polynomial extrapolation, by using $m+1$ different noisy expectation values $\Bqty{\ev{A}\pqty{\lambda_ip_\text{global}}}_{i=0}^{m}$ with $1=\lambda_0<\lambda_1<\cdots<\lambda_m$ and considering the polynomial of degree $m$ in $p_\text{global}^{\frac{1}{2}}$ in \cref{eq:expansion_of_ev_under_global_depo}, the estimator of this polynomial extrapolation is given by
\begin{align}
    \ev{A}_\text{est} = \sum_{i=0}^m g_i\ev{A}\pqty{\lambda_i p_\text{global}} + \mathcal{O}\pqty{p_\text{global}^{\frac{m+1}{2}}},\label{eq:constrained_extrapolation}
\end{align}
where
\begin{align}
    g_i \coloneqq \prod_{j\neq i} \frac{\sqrt{\lambda_j}}{\sqrt{\lambda_j}-\sqrt{\lambda_i}} \label{eq:coeff_of_data_eff_ext}.
\end{align}
We dub this extrapolation as data-efficient 1D extrapolation.

\section{Trotter Subspace Expansion}\label{sec:Trotter_subspace_expansion}
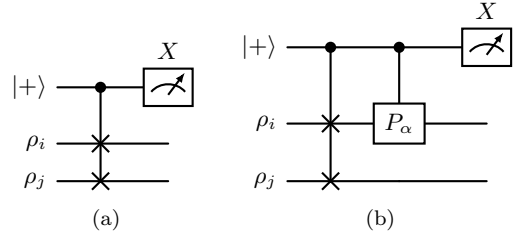
\begin{figure}[tb]
\centering
\subfloat[\label{subfig:swap_test}]{%
    \begin{adjustbox}{valign=b}
        \begin{quantikz}
            \lstick{$\ket{+}$} & \ctrl{1} & \meter{X}\\
            \lstick{$\rho_i$} & \swap{1} &\\
            \lstick{$\rho_j$} & \targX{} &
        \end{quantikz}
    \end{adjustbox}
    } \quad
\subfloat[\label{subfig:circ_for_hij}]{%
    \begin{adjustbox}{valign=b}
        \begin{quantikz}
            \lstick{$\ket{+}$} & \ctrl{1} & \ctrl{1} & \meter{X}\\
            \lstick{$\rho_i$} & \swap{1} & \gate{P_\alpha} &\\
            \lstick{$\rho_j$} & \targX{} & &
        \end{quantikz}
    \end{adjustbox}
    }
        \caption{
        Quantum circuits using the standard purification \cite{koczorDominantEigenvectorNoisy2021,koczorExponentialErrorSuppression2021,hugginsVirtualDistillationQuantum2021,yoshiokaGeneralizedQuantumSubspace2022} for evaluating the expectation value of an observable $P_\alpha$ for the ansatz state of the Trotter subspace $\rho_\text{QEM}$ shown in \cref{eq:Trotter_ansatz}.
        (a): Swap test circuit to evaluate $\text{Tr}\pqty{\rho_i \rho_j}$.
        This circuit allows us to evaluate $\text{Tr}\pqty{\rho_i \rho_j}$ by measuring the Pauli $X$ of the most upper line.
        (b): Quanutm circuit to evaluate $\text{Tr}\pqty{\rho_i \rho_jP_\alpha}$ by measuring the Pauli $X$, where $A=\sum_\alpha c_\alpha P_\alpha$ and $P_\alpha$ is a Pauli operator.
        }
    \label{fig:qc_for_evaluating_of_ev_of_observarble}
\end{figure}
\begin{figure}[tb]
\centering
\subfloat[\label{subfig:dual_se_intermediate}]{%
    \begin{adjustbox}{valign=b}
        \begin{quantikz}
            \lstick{$\ket{\bm{0}}$} & \gate{\mathcal{E}\pqty{p_i,M_i}} &
            \meter{P_\alpha} &\gate{\mathcal{E}_\text{rev}\pqty{p_j,M_j}} &  \rstick{$\ketbra{\bm{0}}{\bm{0}}$}
        \end{quantikz}
    \end{adjustbox}
    } \quad
\subfloat[\label{subfig:dual_se_indirect}]{%
    \begin{adjustbox}{valign=b}
        \begin{quantikz}
            \lstick{$\ket{+}$} & & \ctrl{1} & & \meter{X/Y}\\
            \lstick{$\ket{\bm{0}}$} & \gate{\mathcal{E}\pqty{p_i,M_i}} & \gate{P_\alpha} & \gate{\mathcal{E}_\text{rev}\pqty{p_j, M_j}} & \rstick{$\ketbra{\bm{0}}{\bm{0}}$}
        \end{quantikz}
    \end{adjustbox}
    }
        \caption{
        Quantum circuits using the dual state purification \cite{huoDualstatePurificationPractical2022,caiResourceefficientPurificationbasedQuantum2021,yangResourceefficientGeneralizedQuantum2025} (a) and (b) for evaluating the expectation value of an observable $P_\alpha$ for the ansatz state of the Trotter subspace shown in \cref{eq:dual1,eq:dual2}, respectively.
        }
    \label{fig:dual_qc_for_evaluating_of_ev_of_observarble}
\end{figure}
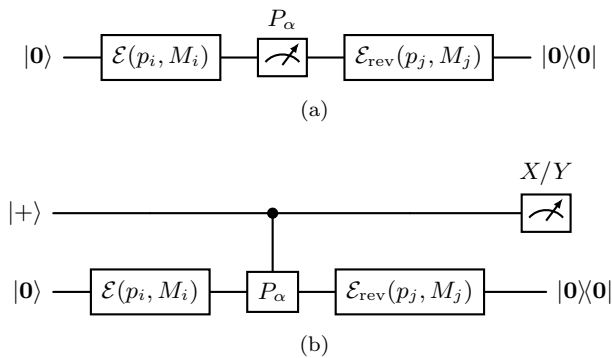
Here, we propose Trotter subspace expansion that robustly mitigates both algorithmic and physical errors more than the proposed data-efficient extrapolation method at the cost of measurements.
The Trotter subspace expansion is the combination of the proposed data-efficient extrapolation method and the purification methods, which certifies the physicality of the estimated result (See \cref{sec:general_remark_of_ext}).

We prepare the $n'+1$ output states of noisy Trotterized circuit $\sigma_i = \rho\pqty{p_i, M_i}$ $\pqty{i=0,\ldots,n'}$,
where $p_i=\lambda_ip_\text{global}$ with $1=\lambda_0<\lambda_1<\cdots<\lambda_{n'}$ and $M_i$ are the error rates of physical noise and the Trotter number of the noisy Trotterized circuit, respectively.
$M_i$ is determined by \cref{eq:opt_Trotter_num_depends_on_physical_error_rate} with a given constant factor $c$.
Then, we virtually construct the following state:
\begin{align}
\rho_\text{TS} &=\sum_{i=0}^{n'} g_i \rho_i, \\
\rho_\text{QEM} &= \frac{\rho_\text{TS}^2}{\text{Tr}(\rho_\text{TS}^2)},
\label{eq:Trotter_ansatz}
\end{align}
where $\rho_i \coloneqq \rho\pqty{p_i,M_i}$ and $g_i$ is determined by \cref{eq:coeff_of_data_eff_ext}.
The ansatz state $\rho_\text{QEM}$ satisfies $\text{Tr}\pqty{\rho_\text{QEM}} = 1$ and $\rho \geq 0$, which ensures that $\rho_\text{QEM}$ is physical.
Then, the corresponding error-mitigated expectation value of $A$ is given by 
\begin{align}
    \ev{A}_\text{est} = 
    \frac{\sum_{i,j=0}^{n'} g_i g_j \text{Tr}\pqty{\rho_i\rho_j A}}{\sum_{i,j=0}^{n'} g_i g_j \text{Tr}\pqty{\rho_i\rho_j}}. \label{eq:estimator_of_trotter_subspace_expansion}
\end{align}
To obtain each term constituting $\ev{A}_\text{est}$, i.e., $\text{Tr}\pqty{\rho_i\rho_j A}$ and  $\text{Tr}\pqty{\rho_i\rho_j}$ in \cref{eq:estimator_of_trotter_subspace_expansion},
we use a modified quantum circuit of purification-based QEM methods. 
The quantum circuits for Trotter subspace expansion for copy-based purification are shown in \cref{fig:qc_for_evaluating_of_ev_of_observarble} \cite{yoshiokaGeneralizedQuantumSubspace2022} (see \cref{sec:explanation_of_circuits} for more details). 

In summary, we can perform the Trotter subspace expansion in the following procedure:
\begin{enumerate}
    \item Choose the $n'+1$ parameter pairs $\pqty{p_i, M_i}$, composed of the physical error rate $p_i$ and Trotter number $M_i$, where $M_i$ is determined by \cref{eq:opt_Trotter_num_depends_on_physical_error_rate}.
    Then, prepare the corresponding noisy Trotterized quantum circuit $\mathcal{C}\pqty{p_i, M_i}$.
    \item For $i,j=0,\ldots,n'$, repeat the following operations.
    \begin{enumerate}
        \item Using the quantum circuits $\mathcal{C}\pqty{p_i, M_i}$ and $\mathcal{C}\pqty{p_j, M_j}$, prepare the output states $\rho_i\coloneqq\rho\pqty{p_i, M_i}$ and $\rho_j$ as the initial states for the quantum circuits shown in \cref{fig:qc_for_evaluating_of_ev_of_observarble}.
        \item Obtain $\text{Tr}\pqty{\rho_i\rho_j}$ and $\text{Tr}\pqty{\rho_i\rho_jA}$ using the quantum circuits shown in \cref{subfig:swap_test,subfig:circ_for_hij}.
    \end{enumerate}
    \item Calculate the coefficients $g_i$ by \cref{eq:coeff_of_data_eff_ext}.
    \item Calculate the denominator and the numerator in \cref{eq:estimator_of_trotter_subspace_expansion} and then output the estimator $\ev{A}_\text{est}$.
\end{enumerate}

To circumvent the use of copies in our method, we can use a dual-state purification circuit~\cite{huoDualstatePurificationPractical2022,caiResourceefficientPurificationbasedQuantum2021,yangResourceefficientGeneralizedQuantum2025}. By the dual-state purification, we can compute the expectation values for following error-mitigated states:

\begin{align}
\rho_\text{Dual}^{(1)} = \frac{\rho_\text{TS} \bar{\rho}_\text{TS}+ \bar{\rho}_\text{TS} \rho_\text{TS}}{2 \text{Tr} (\rho_\text{TS} \bar{\rho}_\text{TS})}
\label{eq:dual1}
\end{align}

or 

\begin{align}
\rho_\text{Dual}^{(2)} = \frac{\rho_\text{TS} \bar{\rho}_\text{TS}}{\mathrm{Tr} (\rho_\text{TS} \bar{\rho}_\text{TS})}
\label{eq:dual2}
\end{align}
with $\bar{\rho}_\text{TS}= \sum_i g_i \bar{\rho}_i$, where $\bar{\rho}_i\coloneqq\bar{\rho}(p_i, M_i)$.
\Cref{fig:dual_qc_for_evaluating_of_ev_of_observarble} shows the quantum circuits for evaluating each term comprising the expectation values for the error-mitigated state in \cref{eq:dual1,eq:dual2} \cite{yangResourceefficientGeneralizedQuantum2025}.
For $\rho_\text{Dual}^{(1)}$, we need to compute  $\text{Tr}\pqty{\rho_i\bar{\rho}_j}$ and $\frac{1}{2}\text{Tr}\pqty{\pqty{\rho_i\bar{\rho}_j+\bar{\rho}_i\rho_j} A}$.
\Cref{subfig:dual_se_intermediate,subfig:dual_se_indirect} shows the quantum circuit for computing those terms for intermediate measurement and indirect measurement methods.
For $\rho_\text{Dual}^{(2)}$, we need to measure  $\text{Tr}\pqty{\rho_i\bar{\rho}_j A}$,
which can be evaluated by measuring the Pauli $Y$ operator in the indirect measurement method and computing $\ev{X}-i \ev{Y}$ (See \cref{sec:explanation_of_circuits} for more details).
Because the dual state $\bar{\rho}$ well-approximate the original state $\rho$ for a deep circuit under the stochastic Pauli noise \cite{yangResourceefficientGeneralizedQuantum2025},
both $\rho_\text{Dual}^{(1)}$ and $\rho_\text{Dual}^{(2)}$ can be employed as a good approximation of the purified state $\rho_\text{QEM}$ in \cref{eq:Trotter_ansatz} in this case. 



\section{Numerical Simulation}\label{sec:numerical_sim}

\begin{figure}
    \centering
    \includegraphics[width=\linewidth]{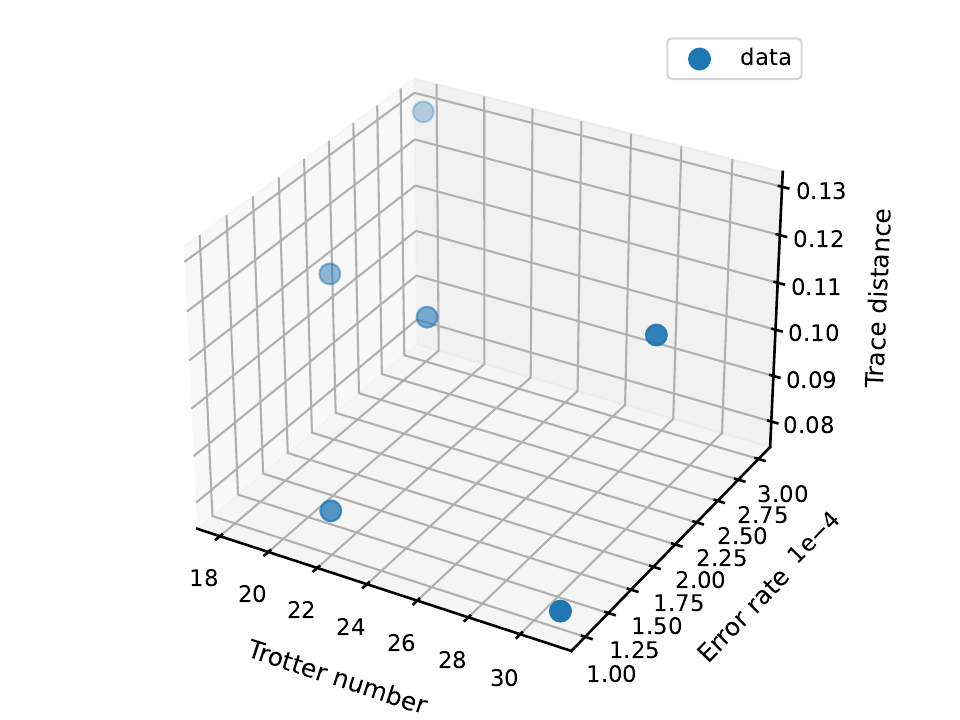}
    \caption{Trace distance between an output state of a noisy Trotterized quantum circuit used for our numerical simulation and the exact state.
    In our numerical simulation, we set the total simulation time as $t=1$ and the physical error rates as $p_1=1.0 \times 10^{-5}$.
    We use the output states such that the corresponding quantum circuits are parametrized by $\pqty{p_2, M} = \pqty{2\times 10^{-4}, 18}$, $\pqty{3\times 10^{-4}, 18}$, $\pqty{1\times 10^{-4}, 22}$, $\pqty{2\times 10^{-4}, 22}$, $\pqty{1\times 10^{-4}, 31}$, and $\pqty{2\times 10^{-4}, 31}$.
    The raw data and the VD use the output state where $\pqty{p_2, M} = \pqty{1\times 10^{-4}, 31}$, whose trace distance is the smallest.
    The previous methods use all of the output states, whereas our proposed data-efficient extrapolation uses only the three output states where $\pqty{p_2,M} = \pqty{3\times 10^{-4}, 18}$, $\pqty{2 \times 10^{-4}, 22}$, and $\pqty{1 \times 10^{-4}, 18}$, satisfying \cref{eq:opt_Trotter_num_depends_on_physical_error_rate}.
    Our proposed Trotter subspace expansion also uses all of the output states, but the number of the required circuits to calculate its estimator is more than those of the previous methods.
    }
    \label{fig:data_set}
\end{figure}
\begin{figure}
    \centering
    \includegraphics[width=\linewidth]{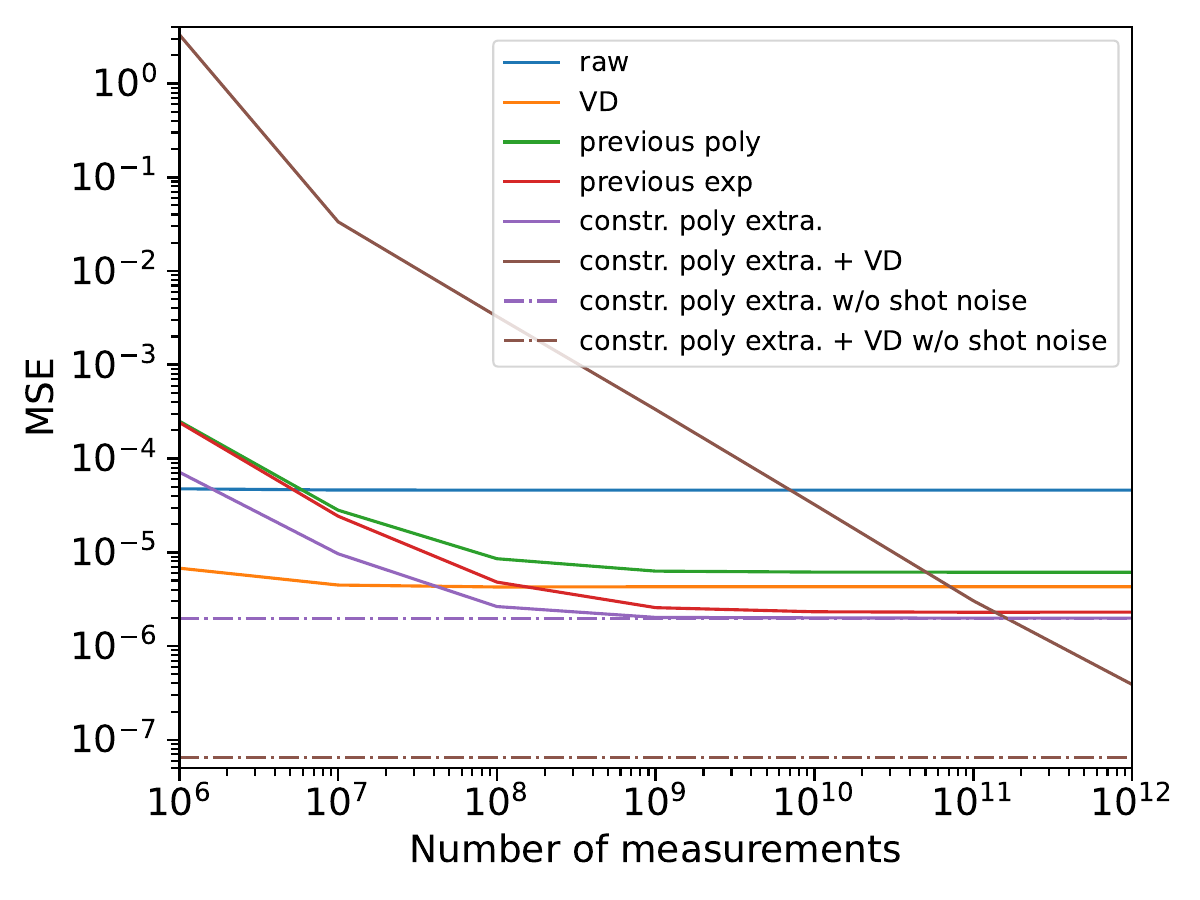}
    \caption{
    Benchmark of QEM methods depending on the number of measurements.
    We estimate the expectation value of an observable $X_1$ for a time-evolved state of the 1D transverse Ising model and calculate the mean squared error (MSE) as a function of the number of measurements.
    The blue, orange, green, red, purple, and brown solid lines stand for the statistics considering shot noise from raw data, VD, the previous method using the polynomial extrapolation in terms of physical noise, the previous method using the exponential extrapolation in terms of physical noise, our proposed data-efficient extrapolation, and the combination of the data-efficient extrapolation with VD, respectively.
    The purple and brown dashed-dotted lines represent the statistics without shot noise from our proposed data-efficient extrapolation and the combination of the data-efficient extrapolation with VD, respectively.
    }
    \label{fig:mse}
\end{figure}
In this section, we present numerical simulations of the Trotter subspace expansion applied to the one-dimensional transverse Ising model.
We initialize the system in the zero state $\ket{0\cdots 0}$ and simulate the time evolution under the Hamiltonian
\begin{align}
    H_\text{TFI} = - \sum_{i=1}^n Z_i Z_{i+1} + \sum_{i=1}^n X_i, 
\end{align}
where $n$ denotes the number of qubits and we set $Z_{n+1} = Z_1$.
We approximate the time evolution under this Hamiltonian by the first-order Trotterization:
\begin{align}
    U_\text{Trotter} = \Bqty{\exp\pqty{i\sum_{i=1}^n Z_i Z_{i+1}\frac{t}{M}}\exp\pqty{-i\sum_{i=1}^n X_i \frac{t}{M}}}^M.
\end{align}
In our simulation, we set $n=10$ and $t=1$, and we assume that there is single- or two-qubit depolarizing noise after a single- or two-qubit gate, respectively.
The single-qubit depolarizing noise that acts on qubit $k$ is defined as
\begin{align}
    \mathcal{N}_\text{1-qubit depo $k$}\pqty{\rho} = \pqty{1-p_1} \rho + \frac{p_1}{2}I_k \otimes \Tr_k\pqty{\rho},
\end{align}
and the two-qubit depolarizing noise that acts on qubit $k$ and qubit $l$ is defined as
\begin{align}
    \mathcal{N}_\text{2-qubit depo $k$, $l$}\pqty{\rho} = \pqty{1-p_2} \rho + \frac{p_2}{4}I_{k,l} \otimes \Tr_{k,l}\pqty{\rho}.
\end{align}
Here, $\rho$ is an $n$-qubit state, and $I_k$ and $I_{k,l}$ are the identity channel for qubit $k$ and that for qubit $k$ and qubit $l$, respectively.
We set the physical error rates as $p_1=1.0 \times 10^{-5}$ and $p_2\in\Bqty{1.0\times 10^{-4}, 2.0\times 10^{-4}, 3.0\times 10^{-4}}$, respectively.
By using \cref{eq:opt_Trotter_num_depends_on_physical_error_rate} with $p_\text{global}=np_2$ and $c=1$, we determine the Trotter number to be $M\in\pqty{18, 22, 31}$.

We benchmark the finite shot noise effect for QEM methods, including our estimators.
We compare our proposed two methods, the 1D data-efficient extrapolation \cref{eq:constrained_extrapolation} and the Trotter subspace expansion \cref{eq:estimator_of_trotter_subspace_expansion}, with other QEM methods such as virtual distillation (VD) \cite{koczorExponentialErrorSuppression2021,hugginsVirtualDistillationQuantum2021} and the previous methods, i.e., the sequential physical and Trotter error extrapolation using a physical polynomial or exponential extrapolation \cite{endoMitigatingAlgorithmicErrors2019}.
Note that we do not consider quantum noise in the purification circuits for the VD and the Trotter subspace expansion shown in \cref{fig:qc_for_evaluating_of_ev_of_observarble} for simplicity.

In \cref{fig:data_set}, we plot the trace distance between the exact time-evolved state and an output state of a noisy Trotterized quantum circuit used for our numerical simulation.
We use the output states such that the corresponding quantum circuits are parametrized by $\pqty{p_2, M} = \pqty{2\times 10^{-4}, 18}$, $\pqty{3\times 10^{-4}, 18}$, $\pqty{1\times 10^{-4}, 22}$, $\pqty{2\times 10^{-4}, 22}$, $\pqty{1\times 10^{-4}, 31}$, and $\pqty{2\times 10^{-4}, 31}$.
The raw data and the VD use the output state where $\pqty{p_2, M} = \pqty{1\times 10^{-4}, 31}$, whose trace distance is the smallest.
The previous methods use all of the output states, whereas our proposed data-efficient extrapolation uses only the three output states where $\pqty{p_2,M} = \pqty{3\times 10^{-4}, 18}$, $\pqty{2 \times 10^{-4}, 22}$, and $\pqty{1 \times 10^{-4}, 18}$, satisfying \cref{eq:opt_Trotter_num_depends_on_physical_error_rate}.
Our proposed Trotter subspace expansion also uses all of the output states, but the number of the required circuits to calculate its estimator is more than those of the previous methods.

To see the effect of the finite number of measurements, we calculate the mean squared error (MSE) of each estimator:
\begin{align}
    \text{MSE} &= \pqty{\ev{A}_\text{exact} - \ev{A}_\text{est}}^2.
\end{align}
We plot the MSE of each estimator as a function of the number of measurements $N$ shown in \cref{fig:mse}.

From \cref{fig:mse}, we find that the MSE of the data-efficient extrapolation is the least in the QEM methods if $10^8\leq N \leq 10^{11}$.
At $10^8$ measurements, we see that the MSE of the data-efficient extrapolation is 17.5 times smaller than the noisy result, 1.6 times smaller than the VD, and 1.8 times smaller than the previous method using the exponential extrapolation in terms of physical noise.
Also, the converged MSE of the data-efficient extrapolation is 23 times smaller than the noisy result, 2.2 times smaller than the VD, and 1.2 times smaller than the previous method using the exponential extrapolation in terms of physical noise.
Thus, the data-efficient extrapolation would be more helpful than the previous methods and the VD.

As for the Trotter subspace expansion, we see that, if we use at least $10^{12}$ measurements, the MSE of the Trotter subspace expansion is the least among the QEM methods.
Also, the converged MSE of the Trotter subspace expansion is 710 times smaller than the raw data,
and 31 times smaller than the data-efficient extrapolation.
Thus, if we can execute many quantum computers in parallel, the Trotter subspace expansion would be more useful than the other QEM methods.

\section{Conclusion}\label{sec:conclusion}
We have presented two error mitigation methods for noisy Trotterized quantum circuits that suppress both physical and algorithmic errors: the data-efficient 1D extrapolation and the Trotter subspace expansion.
The data-efficient 1D extrapolation, which uses the output states parametrized by only one parameter: an error rate of physical noise.
The key point of this method is that it uses the expectation values satisfying the scaling of the optimal Trotter number with respect to a physical error rate, thereby mitigating both algorithmic and physical errors with fewer expectation values than the previous methods \cite{endoMitigatingAlgorithmicErrors2019}.
Then, we proposed the Trotter subspace expansion, i.e., the combination of the data-efficient 1D extrapolation and the VD.
In addition, we have discussed how to implement the Trotter subspace expansion with fewer copies of quantum states by using dual-state purification.

The data-efficient 1D extrapolation is better than the previous methods in terms of the MSE for almost all the measurement shots in our numerical simulation.
The converged MSE of the data-efficient extrapolation is 1.2 times smaller than the previous method using the physical error exponential extrapolation.
We expect that the twirling, such as a symmetry Clifford twirling \cite{tsubouchiSymmetricCliffordTwirling2024}, approximates the quantum noise as the global depolarizing noise, making our proposed extrapolation method more effective.

Moreover, we have numerically seen that the Trotter subspace expansion can be more accurate than the previous methods if we use at least $10^{12}$ measurements.
Its converged MSE is 710 times smaller than the raw data and 31 times smaller than the data-efficient extrapolation.
Thus, the Trotter subspace expansion would be more useful than the other QEM methods if we can execute many quantum computers in parallel.

Our work leaves several open questions.
Although we have considered the first-order deterministic Trotterization, it can be extended to the higher-order Trotterization \cite{suzukiGeneralizedTrotterFormula1976,watsonExponentiallyReducedCircuit2024} or randomized Trotterization \cite{campbellRandomCompilerFast2019,watsonRandomlyCompiledQuantum2025}.
Recently, post-Trotter methods, such as truncated Taylor series \cite{berryExponentialImprovementPrecision2014,berrySimulatingHamiltonianDynamics2015,babbushExponentiallyMorePrecise2016} and quantization \cite{lowHamiltonianSimulationQubitization2019}, have been proposed, and such methods exponentially improve as a function of the desired accuracy compared to the Trotter ones \cite{yoshiokaHuntingQuantumclassicalCrossover2024}.
We also expect that one could give the extension of our work or invention for the post-Trotter methods. 
It would be interesting to apply our proposed method to the statistical phase estimation \cite{linHeisenbergLimitedGroundStateEnergy2022}, which is one of the most promising algorithms for the early-fault tolerant era. Finally, it may be interesting to seek whether our subspace Trotter expansion is helpful in activating the entanglement for quantum metrology~\cite{trenyiActivationMetrologicallyUseful2024}.

\vspace{3mm}

\emph{Note added---.}  When we are finalizing the manuscript, we became aware of a similar work discussing a one-dimensional extrapolation method for both algorithmic and physical errors~\cite{mohammadipourDirectAnalysisZeroNoise2025}.
While their work considers the continuous time evolution model for a noisy second-order Trotter algorithm and adopts a similar parameter tuning to $p \propto M^{-2}$, they do not argue the reason for the optimality of the choice. 
Also, Xu \textit{et. al.} \cite{xuExponentiallyDecayingQuantum2025} have proved that the physical and algorithmic errors in every noisy Trotter step decay exponentially with Trotter steps by the state-dependent analysis.

\begin{acknowledgments}
This work was supported by JST [Moonshot R\&D] Grant Nos. JPMJMS2061; JST, PRESTO, Grant No. JPMJPR2114, Japan; MEXT Q-LEAP, Grant Nos. JPMXS0120319794 and JPMXS0118068682; JST CREST Grant No. JPMJCR23I4.
\end{acknowledgments}
\bibliography{main}

\begin{thebibliography}{44}%
\makeatletter
\providecommand \@ifxundefined [1]{%
 \@ifx{#1\undefined}
}%
\providecommand \@ifnum [1]{%
 \ifnum #1\expandafter \@firstoftwo
 \else \expandafter \@secondoftwo
 \fi
}%
\providecommand \@ifx [1]{%
 \ifx #1\expandafter \@firstoftwo
 \else \expandafter \@secondoftwo
 \fi
}%
\providecommand \natexlab [1]{#1}%
\providecommand \enquote  [1]{``#1''}%
\providecommand \bibnamefont  [1]{#1}%
\providecommand \bibfnamefont [1]{#1}%
\providecommand \citenamefont [1]{#1}%
\providecommand \href@noop [0]{\@secondoftwo}%
\providecommand \href [0]{\begingroup \@sanitize@url \@href}%
\providecommand \@href[1]{\@@startlink{#1}\@@href}%
\providecommand \@@href[1]{\endgroup#1\@@endlink}%
\providecommand \@sanitize@url [0]{\catcode `\\12\catcode `\$12\catcode
  `\&12\catcode `\#12\catcode `\^12\catcode `\_12\catcode `\%12\relax}%
\providecommand \@@startlink[1]{}%
\providecommand \@@endlink[0]{}%
\providecommand \url  [0]{\begingroup\@sanitize@url \@url }%
\providecommand \@url [1]{\endgroup\@href {#1}{\urlprefix }}%
\providecommand \urlprefix  [0]{URL }%
\providecommand \Eprint [0]{\href }%
\providecommand \doibase [0]{https://doi.org/}%
\providecommand \selectlanguage [0]{\@gobble}%
\providecommand \bibinfo  [0]{\@secondoftwo}%
\providecommand \bibfield  [0]{\@secondoftwo}%
\providecommand \translation [1]{[#1]}%
\providecommand \BibitemOpen [0]{}%
\providecommand \bibitemStop [0]{}%
\providecommand \bibitemNoStop [0]{.\EOS\space}%
\providecommand \EOS [0]{\spacefactor3000\relax}%
\providecommand \BibitemShut  [1]{\csname bibitem#1\endcsname}%
\let\auto@bib@innerbib\@empty
\bibitem [{\citenamefont {Kim}\ \emph {et~al.}(2023)\citenamefont {Kim},
  \citenamefont {Eddins}, \citenamefont {Anand}, \citenamefont {Wei},
  \citenamefont {{van den Berg}}, \citenamefont {Rosenblatt}, \citenamefont
  {Nayfeh}, \citenamefont {Wu}, \citenamefont {Zaletel}, \citenamefont
  {Temme},\ and\ \citenamefont {Kandala}}]{kimEvidenceUtilityQuantum2023}%
  \BibitemOpen
  \bibfield  {author} {\bibinfo {author} {\bibfnamefont {Y.}~\bibnamefont
  {Kim}}, \bibinfo {author} {\bibfnamefont {A.}~\bibnamefont {Eddins}},
  \bibinfo {author} {\bibfnamefont {S.}~\bibnamefont {Anand}}, \bibinfo
  {author} {\bibfnamefont {K.~X.}\ \bibnamefont {Wei}}, \bibinfo {author}
  {\bibfnamefont {E.}~\bibnamefont {{van den Berg}}}, \bibinfo {author}
  {\bibfnamefont {S.}~\bibnamefont {Rosenblatt}}, \bibinfo {author}
  {\bibfnamefont {H.}~\bibnamefont {Nayfeh}}, \bibinfo {author} {\bibfnamefont
  {Y.}~\bibnamefont {Wu}}, \bibinfo {author} {\bibfnamefont {M.}~\bibnamefont
  {Zaletel}}, \bibinfo {author} {\bibfnamefont {K.}~\bibnamefont {Temme}},\
  and\ \bibinfo {author} {\bibfnamefont {A.}~\bibnamefont {Kandala}},\
  }\bibfield  {title} {\bibinfo {title} {Evidence for the utility of quantum
  computing before fault tolerance},\ }\href
  {https://doi.org/10.1038/s41586-023-06096-3} {\bibfield  {journal} {\bibinfo
  {journal} {Nature}\ }\textbf {\bibinfo {volume} {618}},\ \bibinfo {pages}
  {500} (\bibinfo {year} {2023})}\BibitemShut {NoStop}%
\bibitem [{\citenamefont {Childs}\ \emph {et~al.}(2018)\citenamefont {Childs},
  \citenamefont {Maslov}, \citenamefont {Nam}, \citenamefont {Ross},\ and\
  \citenamefont {Su}}]{childsFirstQuantumSimulation2018}%
  \BibitemOpen
  \bibfield  {author} {\bibinfo {author} {\bibfnamefont {A.~M.}\ \bibnamefont
  {Childs}}, \bibinfo {author} {\bibfnamefont {D.}~\bibnamefont {Maslov}},
  \bibinfo {author} {\bibfnamefont {Y.}~\bibnamefont {Nam}}, \bibinfo {author}
  {\bibfnamefont {N.~J.}\ \bibnamefont {Ross}},\ and\ \bibinfo {author}
  {\bibfnamefont {Y.}~\bibnamefont {Su}},\ }\bibfield  {title} {\bibinfo
  {title} {Toward the first quantum simulation with quantum speedup},\ }\href
  {https://doi.org/10.1073/pnas.1801723115} {\bibfield  {journal} {\bibinfo
  {journal} {Proc. Natl. Acad. Sci. U.S.A.}\ }\textbf {\bibinfo {volume}
  {115}},\ \bibinfo {pages} {9456} (\bibinfo {year} {2018})}\BibitemShut
  {NoStop}%
\bibitem [{\citenamefont {Yoshioka}\ \emph {et~al.}(2024)\citenamefont
  {Yoshioka}, \citenamefont {Okubo}, \citenamefont {Suzuki}, \citenamefont
  {Koizumi},\ and\ \citenamefont
  {Mizukami}}]{yoshiokaHuntingQuantumclassicalCrossover2024}%
  \BibitemOpen
  \bibfield  {author} {\bibinfo {author} {\bibfnamefont {N.}~\bibnamefont
  {Yoshioka}}, \bibinfo {author} {\bibfnamefont {T.}~\bibnamefont {Okubo}},
  \bibinfo {author} {\bibfnamefont {Y.}~\bibnamefont {Suzuki}}, \bibinfo
  {author} {\bibfnamefont {Y.}~\bibnamefont {Koizumi}},\ and\ \bibinfo {author}
  {\bibfnamefont {W.}~\bibnamefont {Mizukami}},\ }\bibfield  {title} {\bibinfo
  {title} {Hunting for quantum-classical crossover in condensed matter
  problems},\ }\href {https://doi.org/10.1038/s41534-024-00839-4} {\bibfield
  {journal} {\bibinfo  {journal} {npj Quantum Inf}\ }\textbf {\bibinfo {volume}
  {10}},\ \bibinfo {pages} {1} (\bibinfo {year} {2024})}\BibitemShut {NoStop}%
\bibitem [{\citenamefont {Lloyd}(1996)}]{lloydUniversalQuantumSimulators1996}%
  \BibitemOpen
  \bibfield  {author} {\bibinfo {author} {\bibfnamefont {S.}~\bibnamefont
  {Lloyd}},\ }\bibfield  {title} {\bibinfo {title} {Universal {{Quantum
  Simulators}}},\ }\href {https://doi.org/10.1126/science.273.5278.1073}
  {\bibfield  {journal} {\bibinfo  {journal} {Science}\ }\textbf {\bibinfo
  {volume} {273}},\ \bibinfo {pages} {1073} (\bibinfo {year}
  {1996})}\BibitemShut {NoStop}%
\bibitem [{\citenamefont {Suzuki}(1976)}]{suzukiGeneralizedTrotterFormula1976}%
  \BibitemOpen
  \bibfield  {author} {\bibinfo {author} {\bibfnamefont {M.}~\bibnamefont
  {Suzuki}},\ }\bibfield  {title} {\bibinfo {title} {Generalized {{Trotter}}'s
  formula and systematic approximants of exponential operators and inner
  derivations with applications to many-body problems},\ }\href
  {https://doi.org/10.1007/BF01609348} {\bibfield  {journal} {\bibinfo
  {journal} {Commun.Math. Phys.}\ }\textbf {\bibinfo {volume} {51}},\ \bibinfo
  {pages} {183} (\bibinfo {year} {1976})}\BibitemShut {NoStop}%
\bibitem [{\citenamefont {Campbell}(2019)}]{campbellRandomCompilerFast2019}%
  \BibitemOpen
  \bibfield  {author} {\bibinfo {author} {\bibfnamefont {E.}~\bibnamefont
  {Campbell}},\ }\bibfield  {title} {\bibinfo {title} {Random {{Compiler}} for
  {{Fast Hamiltonian Simulation}}},\ }\href
  {https://doi.org/10.1103/PhysRevLett.123.070503} {\bibfield  {journal}
  {\bibinfo  {journal} {Phys. Rev. Lett.}\ }\textbf {\bibinfo {volume} {123}},\
  \bibinfo {pages} {070503} (\bibinfo {year} {2019})}\BibitemShut {NoStop}%
\bibitem [{\citenamefont {Berry}\ \emph {et~al.}(2014)\citenamefont {Berry},
  \citenamefont {Childs}, \citenamefont {Cleve}, \citenamefont {Kothari},\ and\
  \citenamefont {Somma}}]{berryExponentialImprovementPrecision2014}%
  \BibitemOpen
  \bibfield  {author} {\bibinfo {author} {\bibfnamefont {D.~W.}\ \bibnamefont
  {Berry}}, \bibinfo {author} {\bibfnamefont {A.~M.}\ \bibnamefont {Childs}},
  \bibinfo {author} {\bibfnamefont {R.}~\bibnamefont {Cleve}}, \bibinfo
  {author} {\bibfnamefont {R.}~\bibnamefont {Kothari}},\ and\ \bibinfo {author}
  {\bibfnamefont {R.~D.}\ \bibnamefont {Somma}},\ }\bibfield  {title} {\bibinfo
  {title} {Exponential improvement in precision for simulating sparse
  {{Hamiltonians}}},\ }in\ \href {https://doi.org/10.1145/2591796.2591854}
  {\emph {\bibinfo {booktitle} {Proceedings of the Forty-Sixth Annual {{ACM}}
  Symposium on {{Theory}} of Computing}}},\ \bibinfo {series and number}
  {{{STOC}} '14}\ (\bibinfo  {publisher} {Association for Computing
  Machinery},\ \bibinfo {address} {New York, NY, USA},\ \bibinfo {year}
  {2014})\ pp.\ \bibinfo {pages} {283--292}\BibitemShut {NoStop}%
\bibitem [{\citenamefont {Berry}\ \emph {et~al.}(2015)\citenamefont {Berry},
  \citenamefont {Childs}, \citenamefont {Cleve}, \citenamefont {Kothari},\ and\
  \citenamefont {Somma}}]{berrySimulatingHamiltonianDynamics2015}%
  \BibitemOpen
  \bibfield  {author} {\bibinfo {author} {\bibfnamefont {D.~W.}\ \bibnamefont
  {Berry}}, \bibinfo {author} {\bibfnamefont {A.~M.}\ \bibnamefont {Childs}},
  \bibinfo {author} {\bibfnamefont {R.}~\bibnamefont {Cleve}}, \bibinfo
  {author} {\bibfnamefont {R.}~\bibnamefont {Kothari}},\ and\ \bibinfo {author}
  {\bibfnamefont {R.~D.}\ \bibnamefont {Somma}},\ }\bibfield  {title} {\bibinfo
  {title} {Simulating {{Hamiltonian Dynamics}} with a {{Truncated Taylor
  Series}}},\ }\href {https://doi.org/10.1103/PhysRevLett.114.090502}
  {\bibfield  {journal} {\bibinfo  {journal} {Phys. Rev. Lett.}\ }\textbf
  {\bibinfo {volume} {114}},\ \bibinfo {pages} {090502} (\bibinfo {year}
  {2015})}\BibitemShut {NoStop}%
\bibitem [{\citenamefont {Babbush}\ \emph {et~al.}(2016)\citenamefont
  {Babbush}, \citenamefont {Berry}, \citenamefont {Kivlichan}, \citenamefont
  {Wei}, \citenamefont {Love},\ and\ \citenamefont
  {{Aspuru-Guzik}}}]{babbushExponentiallyMorePrecise2016}%
  \BibitemOpen
  \bibfield  {author} {\bibinfo {author} {\bibfnamefont {R.}~\bibnamefont
  {Babbush}}, \bibinfo {author} {\bibfnamefont {D.~W.}\ \bibnamefont {Berry}},
  \bibinfo {author} {\bibfnamefont {I.~D.}\ \bibnamefont {Kivlichan}}, \bibinfo
  {author} {\bibfnamefont {A.~Y.}\ \bibnamefont {Wei}}, \bibinfo {author}
  {\bibfnamefont {P.~J.}\ \bibnamefont {Love}},\ and\ \bibinfo {author}
  {\bibfnamefont {A.}~\bibnamefont {{Aspuru-Guzik}}},\ }\bibfield  {title}
  {\bibinfo {title} {Exponentially more precise quantum simulation of fermions
  in second quantization},\ }\href
  {https://doi.org/10.1088/1367-2630/18/3/033032} {\bibfield  {journal}
  {\bibinfo  {journal} {New J. Phys.}\ }\textbf {\bibinfo {volume} {18}},\
  \bibinfo {pages} {033032} (\bibinfo {year} {2016})}\BibitemShut {NoStop}%
\bibitem [{\citenamefont {Low}\ and\ \citenamefont
  {Chuang}(2019)}]{lowHamiltonianSimulationQubitization2019}%
  \BibitemOpen
  \bibfield  {author} {\bibinfo {author} {\bibfnamefont {G.~H.}\ \bibnamefont
  {Low}}\ and\ \bibinfo {author} {\bibfnamefont {I.~L.}\ \bibnamefont
  {Chuang}},\ }\bibfield  {title} {\bibinfo {title} {Hamiltonian {{Simulation}}
  by {{Qubitization}}},\ }\href {https://doi.org/10.22331/q-2019-07-12-163}
  {\bibfield  {journal} {\bibinfo  {journal} {Quantum}\ }\textbf {\bibinfo
  {volume} {3}},\ \bibinfo {pages} {163} (\bibinfo {year} {2019})}\BibitemShut
  {NoStop}%
\bibitem [{\citenamefont {Temme}\ \emph {et~al.}(2017)\citenamefont {Temme},
  \citenamefont {Bravyi},\ and\ \citenamefont
  {Gambetta}}]{temmeErrorMitigationShortDepth2017}%
  \BibitemOpen
  \bibfield  {author} {\bibinfo {author} {\bibfnamefont {K.}~\bibnamefont
  {Temme}}, \bibinfo {author} {\bibfnamefont {S.}~\bibnamefont {Bravyi}},\ and\
  \bibinfo {author} {\bibfnamefont {J.~M.}\ \bibnamefont {Gambetta}},\
  }\bibfield  {title} {\bibinfo {title} {Error {{Mitigation}} for {{Short-Depth
  Quantum Circuits}}},\ }\href {https://doi.org/10.1103/PhysRevLett.119.180509}
  {\bibfield  {journal} {\bibinfo  {journal} {Phys. Rev. Lett.}\ }\textbf
  {\bibinfo {volume} {119}},\ \bibinfo {pages} {180509} (\bibinfo {year}
  {2017})}\BibitemShut {NoStop}%
\bibitem [{\citenamefont {Li}\ and\ \citenamefont
  {Benjamin}(2017)}]{liEfficientVariationalQuantum2017}%
  \BibitemOpen
  \bibfield  {author} {\bibinfo {author} {\bibfnamefont {Y.}~\bibnamefont
  {Li}}\ and\ \bibinfo {author} {\bibfnamefont {S.~C.}\ \bibnamefont
  {Benjamin}},\ }\bibfield  {title} {\bibinfo {title} {Efficient {{Variational
  Quantum Simulator Incorporating Active Error Minimization}}},\ }\href
  {https://doi.org/10.1103/PhysRevX.7.021050} {\bibfield  {journal} {\bibinfo
  {journal} {Phys. Rev. X}\ }\textbf {\bibinfo {volume} {7}},\ \bibinfo {pages}
  {021050} (\bibinfo {year} {2017})}\BibitemShut {NoStop}%
\bibitem [{\citenamefont {Endo}\ \emph {et~al.}(2018)\citenamefont {Endo},
  \citenamefont {Benjamin},\ and\ \citenamefont
  {Li}}]{endoPracticalQuantumError2018}%
  \BibitemOpen
  \bibfield  {author} {\bibinfo {author} {\bibfnamefont {S.}~\bibnamefont
  {Endo}}, \bibinfo {author} {\bibfnamefont {S.~C.}\ \bibnamefont {Benjamin}},\
  and\ \bibinfo {author} {\bibfnamefont {Y.}~\bibnamefont {Li}},\ }\bibfield
  {title} {\bibinfo {title} {Practical {{Quantum Error Mitigation}} for
  {{Near-Future Applications}}},\ }\href
  {https://doi.org/10.1103/PhysRevX.8.031027} {\bibfield  {journal} {\bibinfo
  {journal} {Phys. Rev. X}\ }\textbf {\bibinfo {volume} {8}},\ \bibinfo {pages}
  {031027} (\bibinfo {year} {2018})}\BibitemShut {NoStop}%
\bibitem [{\citenamefont {Suzuki}\ \emph {et~al.}(2022)\citenamefont {Suzuki},
  \citenamefont {Endo}, \citenamefont {Fujii},\ and\ \citenamefont
  {Tokunaga}}]{suzukiQuantumErrorMitigation2022}%
  \BibitemOpen
  \bibfield  {author} {\bibinfo {author} {\bibfnamefont {Y.}~\bibnamefont
  {Suzuki}}, \bibinfo {author} {\bibfnamefont {S.}~\bibnamefont {Endo}},
  \bibinfo {author} {\bibfnamefont {K.}~\bibnamefont {Fujii}},\ and\ \bibinfo
  {author} {\bibfnamefont {Y.}~\bibnamefont {Tokunaga}},\ }\bibfield  {title}
  {\bibinfo {title} {Quantum {{Error Mitigation}} as a {{Universal Error
  Reduction Technique}}: {{Applications}} from the {{NISQ}} to the
  {{Fault-Tolerant Quantum Computing Eras}}},\ }\href
  {https://doi.org/10.1103/PRXQuantum.3.010345} {\bibfield  {journal} {\bibinfo
   {journal} {PRX Quantum}\ }\textbf {\bibinfo {volume} {3}},\ \bibinfo {pages}
  {010345} (\bibinfo {year} {2022})}\BibitemShut {NoStop}%
\bibitem [{\citenamefont {Xiong}\ \emph {et~al.}(2020)\citenamefont {Xiong},
  \citenamefont {Chandra}, \citenamefont {Ng},\ and\ \citenamefont
  {Hanzo}}]{xiongSamplingOverheadAnalysis2020}%
  \BibitemOpen
  \bibfield  {author} {\bibinfo {author} {\bibfnamefont {Y.}~\bibnamefont
  {Xiong}}, \bibinfo {author} {\bibfnamefont {D.}~\bibnamefont {Chandra}},
  \bibinfo {author} {\bibfnamefont {S.~X.}\ \bibnamefont {Ng}},\ and\ \bibinfo
  {author} {\bibfnamefont {L.}~\bibnamefont {Hanzo}},\ }\bibfield  {title}
  {\bibinfo {title} {Sampling {{Overhead Analysis}} of {{Quantum Error
  Mitigation}}: {{Uncoded}} vs. {{Coded Systems}}},\ }\href
  {https://doi.org/10.1109/ACCESS.2020.3045016} {\bibfield  {journal} {\bibinfo
   {journal} {IEEE Access}\ }\textbf {\bibinfo {volume} {8}},\ \bibinfo {pages}
  {228967} (\bibinfo {year} {2020})}\BibitemShut {NoStop}%
\bibitem [{\citenamefont {Piveteau}\ \emph {et~al.}(2021)\citenamefont
  {Piveteau}, \citenamefont {Sutter}, \citenamefont {Bravyi}, \citenamefont
  {Gambetta},\ and\ \citenamefont
  {Temme}}]{piveteauErrorMitigationUniversal2021}%
  \BibitemOpen
  \bibfield  {author} {\bibinfo {author} {\bibfnamefont {C.}~\bibnamefont
  {Piveteau}}, \bibinfo {author} {\bibfnamefont {D.}~\bibnamefont {Sutter}},
  \bibinfo {author} {\bibfnamefont {S.}~\bibnamefont {Bravyi}}, \bibinfo
  {author} {\bibfnamefont {J.~M.}\ \bibnamefont {Gambetta}},\ and\ \bibinfo
  {author} {\bibfnamefont {K.}~\bibnamefont {Temme}},\ }\bibfield  {title}
  {\bibinfo {title} {Error {{Mitigation}} for {{Universal Gates}} on {{Encoded
  Qubits}}},\ }\href {https://doi.org/10.1103/PhysRevLett.127.200505}
  {\bibfield  {journal} {\bibinfo  {journal} {Phys. Rev. Lett.}\ }\textbf
  {\bibinfo {volume} {127}},\ \bibinfo {pages} {200505} (\bibinfo {year}
  {2021})}\BibitemShut {NoStop}%
\bibitem [{\citenamefont {Lostaglio}\ and\ \citenamefont
  {Ciani}(2021)}]{lostaglioErrorMitigationQuantumAssisted2021}%
  \BibitemOpen
  \bibfield  {author} {\bibinfo {author} {\bibfnamefont {M.}~\bibnamefont
  {Lostaglio}}\ and\ \bibinfo {author} {\bibfnamefont {A.}~\bibnamefont
  {Ciani}},\ }\bibfield  {title} {\bibinfo {title} {Error {{Mitigation}} and
  {{Quantum-Assisted Simulation}} in the {{Error Corrected Regime}}},\ }\href
  {https://doi.org/10.1103/PhysRevLett.127.200506} {\bibfield  {journal}
  {\bibinfo  {journal} {Phys. Rev. Lett.}\ }\textbf {\bibinfo {volume} {127}},\
  \bibinfo {pages} {200506} (\bibinfo {year} {2021})}\BibitemShut {NoStop}%
\bibitem [{\citenamefont {Wahl}\ \emph {et~al.}(2023)\citenamefont {Wahl},
  \citenamefont {Mari}, \citenamefont {Shammah}, \citenamefont {Zeng},\ and\
  \citenamefont {Ravi}}]{wahlZeroNoiseExtrapolation2023}%
  \BibitemOpen
  \bibfield  {author} {\bibinfo {author} {\bibfnamefont {M.~A.}\ \bibnamefont
  {Wahl}}, \bibinfo {author} {\bibfnamefont {A.}~\bibnamefont {Mari}}, \bibinfo
  {author} {\bibfnamefont {N.}~\bibnamefont {Shammah}}, \bibinfo {author}
  {\bibfnamefont {W.~J.}\ \bibnamefont {Zeng}},\ and\ \bibinfo {author}
  {\bibfnamefont {G.~S.}\ \bibnamefont {Ravi}},\ }\bibfield  {title} {\bibinfo
  {title} {Zero {{Noise Extrapolation}} on {{Logical Qubits}} by {{Scaling}}
  the {{Error Correction Code Distance}}},\ }in\ \href
  {https://doi.org/10.1109/QCE57702.2023.00103} {\emph {\bibinfo {booktitle}
  {2023 {{IEEE International Conference}} on {{Quantum Computing}} and
  {{Engineering}} ({{QCE}})}}}\ (\bibinfo  {publisher} {IEEE},\ \bibinfo
  {address} {Bellevue, WA, USA},\ \bibinfo {year} {2023})\ pp.\ \bibinfo
  {pages} {888--897}\BibitemShut {NoStop}%
\bibitem [{\citenamefont {Endo}\ \emph {et~al.}(2019)\citenamefont {Endo},
  \citenamefont {Zhao}, \citenamefont {Li}, \citenamefont {Benjamin},\ and\
  \citenamefont {Yuan}}]{endoMitigatingAlgorithmicErrors2019}%
  \BibitemOpen
  \bibfield  {author} {\bibinfo {author} {\bibfnamefont {S.}~\bibnamefont
  {Endo}}, \bibinfo {author} {\bibfnamefont {Q.}~\bibnamefont {Zhao}}, \bibinfo
  {author} {\bibfnamefont {Y.}~\bibnamefont {Li}}, \bibinfo {author}
  {\bibfnamefont {S.}~\bibnamefont {Benjamin}},\ and\ \bibinfo {author}
  {\bibfnamefont {X.}~\bibnamefont {Yuan}},\ }\bibfield  {title} {\bibinfo
  {title} {Mitigating algorithmic errors in a {{Hamiltonian}} simulation},\
  }\href {https://doi.org/10.1103/PhysRevA.99.012334} {\bibfield  {journal}
  {\bibinfo  {journal} {Phys. Rev. A}\ }\textbf {\bibinfo {volume} {99}},\
  \bibinfo {pages} {012334} (\bibinfo {year} {2019})}\BibitemShut {NoStop}%
\bibitem [{\citenamefont
  {Koczor}(2021{\natexlab{a}})}]{koczorExponentialErrorSuppression2021}%
  \BibitemOpen
  \bibfield  {author} {\bibinfo {author} {\bibfnamefont {B.}~\bibnamefont
  {Koczor}},\ }\bibfield  {title} {\bibinfo {title} {Exponential {{Error
  Suppression}} for {{Near-Term Quantum Devices}}},\ }\href
  {https://doi.org/10.1103/PhysRevX.11.031057} {\bibfield  {journal} {\bibinfo
  {journal} {Phys. Rev. X}\ }\textbf {\bibinfo {volume} {11}},\ \bibinfo
  {pages} {031057} (\bibinfo {year} {2021}{\natexlab{a}})}\BibitemShut
  {NoStop}%
\bibitem [{\citenamefont
  {Koczor}(2021{\natexlab{b}})}]{koczorDominantEigenvectorNoisy2021}%
  \BibitemOpen
  \bibfield  {author} {\bibinfo {author} {\bibfnamefont {B.}~\bibnamefont
  {Koczor}},\ }\bibfield  {title} {\bibinfo {title} {The dominant eigenvector
  of a noisy quantum state},\ }\href {https://doi.org/10.1088/1367-2630/ac37ae}
  {\bibfield  {journal} {\bibinfo  {journal} {New J. Phys.}\ }\textbf {\bibinfo
  {volume} {23}},\ \bibinfo {pages} {123047} (\bibinfo {year}
  {2021}{\natexlab{b}})}\BibitemShut {NoStop}%
\bibitem [{\citenamefont {Huggins}\ \emph {et~al.}(2021)\citenamefont
  {Huggins}, \citenamefont {McArdle}, \citenamefont {O'Brien}, \citenamefont
  {Lee}, \citenamefont {Rubin}, \citenamefont {Boixo}, \citenamefont {Whaley},
  \citenamefont {Babbush},\ and\ \citenamefont
  {McClean}}]{hugginsVirtualDistillationQuantum2021}%
  \BibitemOpen
  \bibfield  {author} {\bibinfo {author} {\bibfnamefont {W.~J.}\ \bibnamefont
  {Huggins}}, \bibinfo {author} {\bibfnamefont {S.}~\bibnamefont {McArdle}},
  \bibinfo {author} {\bibfnamefont {T.~E.}\ \bibnamefont {O'Brien}}, \bibinfo
  {author} {\bibfnamefont {J.}~\bibnamefont {Lee}}, \bibinfo {author}
  {\bibfnamefont {N.~C.}\ \bibnamefont {Rubin}}, \bibinfo {author}
  {\bibfnamefont {S.}~\bibnamefont {Boixo}}, \bibinfo {author} {\bibfnamefont
  {K.~B.}\ \bibnamefont {Whaley}}, \bibinfo {author} {\bibfnamefont
  {R.}~\bibnamefont {Babbush}},\ and\ \bibinfo {author} {\bibfnamefont {J.~R.}\
  \bibnamefont {McClean}},\ }\bibfield  {title} {\bibinfo {title} {Virtual
  {{Distillation}} for {{Quantum Error Mitigation}}},\ }\href
  {https://doi.org/10.1103/PhysRevX.11.041036} {\bibfield  {journal} {\bibinfo
  {journal} {Phys. Rev. X}\ }\textbf {\bibinfo {volume} {11}},\ \bibinfo
  {pages} {041036} (\bibinfo {year} {2021})}\BibitemShut {NoStop}%
\bibitem [{\citenamefont
  {Cai}(2021{\natexlab{a}})}]{caiResourceefficientPurificationbasedQuantum2021}%
  \BibitemOpen
  \bibfield  {author} {\bibinfo {author} {\bibfnamefont {Z.}~\bibnamefont
  {Cai}},\ }\href {https://doi.org/10.48550/arXiv.2107.07279} {\bibinfo {title}
  {Resource-efficient {{Purification-based Quantum Error Mitigation}}}}
  (\bibinfo {year} {2021}{\natexlab{a}}),\ \Eprint
  {https://arxiv.org/abs/2107.07279} {arXiv:2107.07279} \BibitemShut {NoStop}%
\bibitem [{\citenamefont {Poulin}\ and\ \citenamefont
  {Wocjan}(2009)}]{poulinSamplingThermalQuantum2009}%
  \BibitemOpen
  \bibfield  {author} {\bibinfo {author} {\bibfnamefont {D.}~\bibnamefont
  {Poulin}}\ and\ \bibinfo {author} {\bibfnamefont {P.}~\bibnamefont
  {Wocjan}},\ }\bibfield  {title} {\bibinfo {title} {Sampling from the
  {{Thermal Quantum Gibbs State}} and {{Evaluating Partition Functions}} with a
  {{Quantum Computer}}},\ }\href
  {https://doi.org/10.1103/PhysRevLett.103.220502} {\bibfield  {journal}
  {\bibinfo  {journal} {Phys. Rev. Lett.}\ }\textbf {\bibinfo {volume} {103}},\
  \bibinfo {pages} {220502} (\bibinfo {year} {2009})}\BibitemShut {NoStop}%
\bibitem [{\citenamefont {Nielsen}\ and\ \citenamefont
  {Chuang}(2010)}]{nielsenQuantumComputationQuantum2010}%
  \BibitemOpen
  \bibfield  {author} {\bibinfo {author} {\bibfnamefont {M.~A.}\ \bibnamefont
  {Nielsen}}\ and\ \bibinfo {author} {\bibfnamefont {I.~L.}\ \bibnamefont
  {Chuang}},\ }\href@noop {} {\emph {\bibinfo {title} {Quantum Computation and
  Quantum Information}}},\ \bibinfo {edition} {10th}\ ed.\ (\bibinfo
  {publisher} {Cambridge university press},\ \bibinfo {address} {Cambridge},\
  \bibinfo {year} {2010})\BibitemShut {NoStop}%
\bibitem [{\citenamefont {Lin}\ and\ \citenamefont
  {Tong}(2022)}]{linHeisenbergLimitedGroundStateEnergy2022}%
  \BibitemOpen
  \bibfield  {author} {\bibinfo {author} {\bibfnamefont {L.}~\bibnamefont
  {Lin}}\ and\ \bibinfo {author} {\bibfnamefont {Y.}~\bibnamefont {Tong}},\
  }\bibfield  {title} {\bibinfo {title} {Heisenberg-{{Limited Ground-State
  Energy Estimation}} for {{Early Fault-Tolerant Quantum Computers}}},\ }\href
  {https://doi.org/10.1103/PRXQuantum.3.010318} {\bibfield  {journal} {\bibinfo
   {journal} {PRX Quantum}\ }\textbf {\bibinfo {volume} {3}},\ \bibinfo {pages}
  {010318} (\bibinfo {year} {2022})}\BibitemShut {NoStop}%
\bibitem [{\citenamefont {Knee}\ and\ \citenamefont
  {Munro}(2015)}]{kneeOptimalTrotterizationUniversal2015}%
  \BibitemOpen
  \bibfield  {author} {\bibinfo {author} {\bibfnamefont {G.~C.}\ \bibnamefont
  {Knee}}\ and\ \bibinfo {author} {\bibfnamefont {W.~J.}\ \bibnamefont
  {Munro}},\ }\bibfield  {title} {\bibinfo {title} {Optimal {{Trotterization}}
  in universal quantum simulators under faulty control},\ }\bibfield  {journal}
  {\bibinfo  {journal} {Physical Review A}\ }\textbf {\bibinfo {volume} {91}},\
  \href {https://doi.org/10.1103/PhysRevA.91.052327}
  {10.1103/PhysRevA.91.052327} (\bibinfo {year} {2015})\BibitemShut {NoStop}%
\bibitem [{\citenamefont {Gilchrist}\ \emph {et~al.}(2005)\citenamefont
  {Gilchrist}, \citenamefont {Langford},\ and\ \citenamefont
  {Nielsen}}]{gilchristDistanceMeasuresCompare2005}%
  \BibitemOpen
  \bibfield  {author} {\bibinfo {author} {\bibfnamefont {A.}~\bibnamefont
  {Gilchrist}}, \bibinfo {author} {\bibfnamefont {N.~K.}\ \bibnamefont
  {Langford}},\ and\ \bibinfo {author} {\bibfnamefont {M.~A.}\ \bibnamefont
  {Nielsen}},\ }\bibfield  {title} {\bibinfo {title} {Distance measures to
  compare real and ideal quantum processes},\ }\href
  {https://doi.org/10.1103/PhysRevA.71.062310} {\bibfield  {journal} {\bibinfo
  {journal} {Phys. Rev. A}\ }\textbf {\bibinfo {volume} {71}},\ \bibinfo
  {pages} {062310} (\bibinfo {year} {2005})}\BibitemShut {NoStop}%
\bibitem [{\citenamefont {Qin}\ \emph {et~al.}(2023)\citenamefont {Qin},
  \citenamefont {Chen},\ and\ \citenamefont
  {Li}}]{qinErrorStatisticsScalability2023}%
  \BibitemOpen
  \bibfield  {author} {\bibinfo {author} {\bibfnamefont {D.}~\bibnamefont
  {Qin}}, \bibinfo {author} {\bibfnamefont {Y.}~\bibnamefont {Chen}},\ and\
  \bibinfo {author} {\bibfnamefont {Y.}~\bibnamefont {Li}},\ }\bibfield
  {title} {\bibinfo {title} {Error statistics and scalability of quantum error
  mitigation formulas},\ }\href {https://doi.org/10.1038/s41534-023-00707-7}
  {\bibfield  {journal} {\bibinfo  {journal} {npj Quantum Inf}\ }\textbf
  {\bibinfo {volume} {9}},\ \bibinfo {pages} {1} (\bibinfo {year}
  {2023})}\BibitemShut {NoStop}%
\bibitem [{\citenamefont {Tsubouchi}\ \emph {et~al.}(2023)\citenamefont
  {Tsubouchi}, \citenamefont {Sagawa},\ and\ \citenamefont
  {Yoshioka}}]{tsubouchiUniversalCostBound2023}%
  \BibitemOpen
  \bibfield  {author} {\bibinfo {author} {\bibfnamefont {K.}~\bibnamefont
  {Tsubouchi}}, \bibinfo {author} {\bibfnamefont {T.}~\bibnamefont {Sagawa}},\
  and\ \bibinfo {author} {\bibfnamefont {N.}~\bibnamefont {Yoshioka}},\
  }\bibfield  {title} {\bibinfo {title} {Universal {{Cost Bound}} of {{Quantum
  Error Mitigation Based}} on {{Quantum Estimation Theory}}},\ }\href
  {https://doi.org/10.1103/PhysRevLett.131.210601} {\bibfield  {journal}
  {\bibinfo  {journal} {Phys. Rev. Lett.}\ }\textbf {\bibinfo {volume} {131}},\
  \bibinfo {pages} {210601} (\bibinfo {year} {2023})}\BibitemShut {NoStop}%
\bibitem [{\citenamefont {Childs}\ \emph {et~al.}(2021)\citenamefont {Childs},
  \citenamefont {Su}, \citenamefont {Tran}, \citenamefont {Wiebe},\ and\
  \citenamefont {Zhu}}]{childsTheoryTrotterError2021}%
  \BibitemOpen
  \bibfield  {author} {\bibinfo {author} {\bibfnamefont {A.~M.}\ \bibnamefont
  {Childs}}, \bibinfo {author} {\bibfnamefont {Y.}~\bibnamefont {Su}}, \bibinfo
  {author} {\bibfnamefont {M.~C.}\ \bibnamefont {Tran}}, \bibinfo {author}
  {\bibfnamefont {N.}~\bibnamefont {Wiebe}},\ and\ \bibinfo {author}
  {\bibfnamefont {S.}~\bibnamefont {Zhu}},\ }\bibfield  {title} {\bibinfo
  {title} {Theory of {{Trotter Error}} with {{Commutator Scaling}}},\ }\href
  {https://doi.org/10.1103/PhysRevX.11.011020} {\bibfield  {journal} {\bibinfo
  {journal} {Phys. Rev. X}\ }\textbf {\bibinfo {volume} {11}},\ \bibinfo
  {pages} {011020} (\bibinfo {year} {2021})}\BibitemShut {NoStop}%
\bibitem [{\citenamefont {Sanders}\ \emph {et~al.}(2007)\citenamefont
  {Sanders}, \citenamefont {Ahokas}, \citenamefont {Cleve},\ and\ \citenamefont
  {Berry}}]{sandersQuantumAlgorithmsHamiltonian2007}%
  \BibitemOpen
  \bibfield  {author} {\bibinfo {author} {\bibfnamefont {B.}~\bibnamefont
  {Sanders}}, \bibinfo {author} {\bibfnamefont {G.}~\bibnamefont {Ahokas}},
  \bibinfo {author} {\bibfnamefont {R.}~\bibnamefont {Cleve}},\ and\ \bibinfo
  {author} {\bibfnamefont {D.}~\bibnamefont {Berry}},\ }\bibfield  {title}
  {\bibinfo {title} {Quantum {{Algorithms}} for {{Hamiltonian Simulation}}},\
  }in\ \href {https://doi.org/10.1201/9781584889007.ch4} {\emph {\bibinfo
  {booktitle} {Mathematics of {{Quantum Computation}} and {{Quantum
  Technology}}}}},\ Vol.\ \bibinfo {volume} {20074453},\ \bibinfo {editor}
  {edited by\ \bibinfo {editor} {\bibfnamefont {G.}~\bibnamefont {Chen}},
  \bibinfo {editor} {\bibfnamefont {L.}~\bibnamefont {Kauffman}},\ and\
  \bibinfo {editor} {\bibfnamefont {S.}~\bibnamefont {Lomonaco}}}\ (\bibinfo
  {publisher} {{Chapman and Hall/CRC}},\ \bibinfo {year} {2007})\ pp.\ \bibinfo
  {pages} {89--112}\BibitemShut {NoStop}%
\bibitem [{\citenamefont {Yoshioka}\ \emph {et~al.}(2022)\citenamefont
  {Yoshioka}, \citenamefont {Hakoshima}, \citenamefont {Matsuzaki},
  \citenamefont {Tokunaga}, \citenamefont {Suzuki},\ and\ \citenamefont
  {Endo}}]{yoshiokaGeneralizedQuantumSubspace2022}%
  \BibitemOpen
  \bibfield  {author} {\bibinfo {author} {\bibfnamefont {N.}~\bibnamefont
  {Yoshioka}}, \bibinfo {author} {\bibfnamefont {H.}~\bibnamefont {Hakoshima}},
  \bibinfo {author} {\bibfnamefont {Y.}~\bibnamefont {Matsuzaki}}, \bibinfo
  {author} {\bibfnamefont {Y.}~\bibnamefont {Tokunaga}}, \bibinfo {author}
  {\bibfnamefont {Y.}~\bibnamefont {Suzuki}},\ and\ \bibinfo {author}
  {\bibfnamefont {S.}~\bibnamefont {Endo}},\ }\bibfield  {title} {\bibinfo
  {title} {Generalized {{Quantum Subspace Expansion}}},\ }\href
  {https://doi.org/10.1103/PhysRevLett.129.020502} {\bibfield  {journal}
  {\bibinfo  {journal} {Phys. Rev. Lett.}\ }\textbf {\bibinfo {volume} {129}},\
  \bibinfo {pages} {020502} (\bibinfo {year} {2022})}\BibitemShut {NoStop}%
\bibitem [{\citenamefont {Huo}\ and\ \citenamefont
  {Li}(2022)}]{huoDualstatePurificationPractical2022}%
  \BibitemOpen
  \bibfield  {author} {\bibinfo {author} {\bibfnamefont {M.}~\bibnamefont
  {Huo}}\ and\ \bibinfo {author} {\bibfnamefont {Y.}~\bibnamefont {Li}},\
  }\bibfield  {title} {\bibinfo {title} {Dual-state purification for practical
  quantum error mitigation},\ }\href
  {https://doi.org/10.1103/PhysRevA.105.022427} {\bibfield  {journal} {\bibinfo
   {journal} {Phys. Rev. A}\ }\textbf {\bibinfo {volume} {105}},\ \bibinfo
  {pages} {022427} (\bibinfo {year} {2022})}\BibitemShut {NoStop}%
\bibitem [{\citenamefont {Yang}\ \emph {et~al.}(2025)\citenamefont {Yang},
  \citenamefont {Yoshioka}, \citenamefont {Harada}, \citenamefont {Hakkaku},
  \citenamefont {Tokunaga}, \citenamefont {Hakoshima}, \citenamefont
  {Yamamoto},\ and\ \citenamefont
  {Endo}}]{yangResourceefficientGeneralizedQuantum2025}%
  \BibitemOpen
  \bibfield  {author} {\bibinfo {author} {\bibfnamefont {B.}~\bibnamefont
  {Yang}}, \bibinfo {author} {\bibfnamefont {N.}~\bibnamefont {Yoshioka}},
  \bibinfo {author} {\bibfnamefont {H.}~\bibnamefont {Harada}}, \bibinfo
  {author} {\bibfnamefont {S.}~\bibnamefont {Hakkaku}}, \bibinfo {author}
  {\bibfnamefont {Y.}~\bibnamefont {Tokunaga}}, \bibinfo {author}
  {\bibfnamefont {H.}~\bibnamefont {Hakoshima}}, \bibinfo {author}
  {\bibfnamefont {K.}~\bibnamefont {Yamamoto}},\ and\ \bibinfo {author}
  {\bibfnamefont {S.}~\bibnamefont {Endo}},\ }\bibfield  {title} {\bibinfo
  {title} {Resource-efficient generalized quantum subspace expansion},\ }\href
  {https://doi.org/10.1103/PhysRevApplied.23.054021} {\bibfield  {journal}
  {\bibinfo  {journal} {Phys. Rev. Appl.}\ }\textbf {\bibinfo {volume} {23}},\
  \bibinfo {pages} {054021} (\bibinfo {year} {2025})}\BibitemShut {NoStop}%
\bibitem [{\citenamefont {Tsubouchi}\ \emph {et~al.}(2024)\citenamefont
  {Tsubouchi}, \citenamefont {Mitsuhashi}, \citenamefont {Sharma},\ and\
  \citenamefont {Yoshioka}}]{tsubouchiSymmetricCliffordTwirling2024}%
  \BibitemOpen
  \bibfield  {author} {\bibinfo {author} {\bibfnamefont {K.}~\bibnamefont
  {Tsubouchi}}, \bibinfo {author} {\bibfnamefont {Y.}~\bibnamefont
  {Mitsuhashi}}, \bibinfo {author} {\bibfnamefont {K.}~\bibnamefont {Sharma}},\
  and\ \bibinfo {author} {\bibfnamefont {N.}~\bibnamefont {Yoshioka}},\
  }\href@noop {} {\bibinfo {title} {Symmetric {{Clifford}} twirling for
  cost-optimal quantum error mitigation in early {{FTQC}} regime}} (\bibinfo
  {year} {2024}),\ \Eprint {https://arxiv.org/abs/2405.07720}
  {arXiv:2405.07720} \BibitemShut {NoStop}%
\bibitem [{\citenamefont {Watson}\ and\ \citenamefont
  {Watkins}(2024)}]{watsonExponentiallyReducedCircuit2024}%
  \BibitemOpen
  \bibfield  {author} {\bibinfo {author} {\bibfnamefont {J.~D.}\ \bibnamefont
  {Watson}}\ and\ \bibinfo {author} {\bibfnamefont {J.}~\bibnamefont
  {Watkins}},\ }\href {https://doi.org/10.48550/arXiv.2408.14385} {\bibinfo
  {title} {Exponentially {{Reduced Circuit Depths Using Trotter Error
  Mitigation}}}} (\bibinfo {year} {2024}),\ \Eprint
  {https://arxiv.org/abs/2408.14385} {arXiv:2408.14385} \BibitemShut {NoStop}%
\bibitem [{\citenamefont {Watson}(2025)}]{watsonRandomlyCompiledQuantum2025}%
  \BibitemOpen
  \bibfield  {author} {\bibinfo {author} {\bibfnamefont {J.~D.}\ \bibnamefont
  {Watson}},\ }\href {https://doi.org/10.48550/arXiv.2411.04240} {\bibinfo
  {title} {Randomly {{Compiled Quantum Simulation}} with {{Exponentially
  Reduced Circuit Depths}}}} (\bibinfo {year} {2025}),\ \Eprint
  {https://arxiv.org/abs/2411.04240} {arXiv:2411.04240} \BibitemShut {NoStop}%
\bibitem [{\citenamefont {Tr{\'e}nyi}\ \emph {et~al.}(2024)\citenamefont
  {Tr{\'e}nyi}, \citenamefont {Luk{\'a}cs}, \citenamefont {Horodecki},
  \citenamefont {Horodecki}, \citenamefont {V{\'e}rtesi},\ and\ \citenamefont
  {T{\'o}th}}]{trenyiActivationMetrologicallyUseful2024}%
  \BibitemOpen
  \bibfield  {author} {\bibinfo {author} {\bibfnamefont {R.}~\bibnamefont
  {Tr{\'e}nyi}}, \bibinfo {author} {\bibfnamefont {{\'A}.}~\bibnamefont
  {Luk{\'a}cs}}, \bibinfo {author} {\bibfnamefont {P.}~\bibnamefont
  {Horodecki}}, \bibinfo {author} {\bibfnamefont {R.}~\bibnamefont
  {Horodecki}}, \bibinfo {author} {\bibfnamefont {T.}~\bibnamefont
  {V{\'e}rtesi}},\ and\ \bibinfo {author} {\bibfnamefont {G.}~\bibnamefont
  {T{\'o}th}},\ }\bibfield  {title} {\bibinfo {title} {Activation of
  metrologically useful genuine multipartite entanglement},\ }\href
  {https://doi.org/10.1088/1367-2630/ad1e93} {\bibfield  {journal} {\bibinfo
  {journal} {New J. Phys.}\ }\textbf {\bibinfo {volume} {26}},\ \bibinfo
  {pages} {023034} (\bibinfo {year} {2024})}\BibitemShut {NoStop}%
\bibitem [{\citenamefont {Mohammadipour}\ and\ \citenamefont
  {Li}(2025)}]{mohammadipourDirectAnalysisZeroNoise2025}%
  \BibitemOpen
  \bibfield  {author} {\bibinfo {author} {\bibfnamefont {P.}~\bibnamefont
  {Mohammadipour}}\ and\ \bibinfo {author} {\bibfnamefont {X.}~\bibnamefont
  {Li}},\ }\href {https://doi.org/10.48550/arXiv.2502.20673} {\bibinfo {title}
  {Direct {{Analysis}} of {{Zero-Noise Extrapolation}}: {{Polynomial Methods}},
  {{Error Bounds}}, and {{Simultaneous Physical-Algorithmic Error
  Mitigation}}}} (\bibinfo {year} {2025}),\ \Eprint
  {https://arxiv.org/abs/2502.20673} {arXiv:2502.20673} \BibitemShut {NoStop}%
\bibitem [{\citenamefont {Xu}\ \emph {et~al.}(2025)\citenamefont {Xu},
  \citenamefont {Zhao}, \citenamefont {Fan},\ and\ \citenamefont
  {Zhao}}]{xuExponentiallyDecayingQuantum2025}%
  \BibitemOpen
  \bibfield  {author} {\bibinfo {author} {\bibfnamefont {J.}~\bibnamefont
  {Xu}}, \bibinfo {author} {\bibfnamefont {C.}~\bibnamefont {Zhao}}, \bibinfo
  {author} {\bibfnamefont {J.}~\bibnamefont {Fan}},\ and\ \bibinfo {author}
  {\bibfnamefont {Q.}~\bibnamefont {Zhao}},\ }\href
  {https://doi.org/10.48550/arXiv.2504.10247} {\bibinfo {title} {Exponentially
  {{Decaying Quantum Simulation Error}} with {{Noisy Devices}}}} (\bibinfo
  {year} {2025}),\ \Eprint {https://arxiv.org/abs/2504.10247}
  {arXiv:2504.10247} \BibitemShut {NoStop}%
\bibitem [{\citenamefont {Endo}\ \emph {et~al.}(2021)\citenamefont {Endo},
  \citenamefont {Cai}, \citenamefont {Benjamin},\ and\ \citenamefont
  {Yuan}}]{endoHybridQuantumClassicalAlgorithms2021}%
  \BibitemOpen
  \bibfield  {author} {\bibinfo {author} {\bibfnamefont {S.}~\bibnamefont
  {Endo}}, \bibinfo {author} {\bibfnamefont {Z.}~\bibnamefont {Cai}}, \bibinfo
  {author} {\bibfnamefont {S.~C.}\ \bibnamefont {Benjamin}},\ and\ \bibinfo
  {author} {\bibfnamefont {X.}~\bibnamefont {Yuan}},\ }\bibfield  {title}
  {\bibinfo {title} {Hybrid {{Quantum-Classical Algorithms}} and {{Quantum
  Error Mitigation}}},\ }\href {https://doi.org/10.7566/JPSJ.90.032001}
  {\bibfield  {journal} {\bibinfo  {journal} {J. Phys. Soc. Jpn.}\ }\textbf
  {\bibinfo {volume} {90}},\ \bibinfo {pages} {032001} (\bibinfo {year}
  {2021})}\BibitemShut {NoStop}%
\bibitem [{\citenamefont
  {Cai}(2021{\natexlab{b}})}]{caiQuantumErrorMitigation2021}%
  \BibitemOpen
  \bibfield  {author} {\bibinfo {author} {\bibfnamefont {Z.}~\bibnamefont
  {Cai}},\ }\bibfield  {title} {\bibinfo {title} {Quantum {{Error Mitigation}}
  using {{Symmetry Expansion}}},\ }\href
  {https://doi.org/10.22331/q-2021-09-21-548} {\bibfield  {journal} {\bibinfo
  {journal} {Quantum}\ }\textbf {\bibinfo {volume} {5}},\ \bibinfo {pages}
  {548} (\bibinfo {year} {2021}{\natexlab{b}})}\BibitemShut {NoStop}%
\bibitem [{\citenamefont {Kandala}\ \emph {et~al.}(2019)\citenamefont
  {Kandala}, \citenamefont {Temme}, \citenamefont {C{\'o}rcoles}, \citenamefont
  {Mezzacapo}, \citenamefont {Chow},\ and\ \citenamefont
  {Gambetta}}]{kandalaErrorMitigationExtends2019}%
  \BibitemOpen
  \bibfield  {author} {\bibinfo {author} {\bibfnamefont {A.}~\bibnamefont
  {Kandala}}, \bibinfo {author} {\bibfnamefont {K.}~\bibnamefont {Temme}},
  \bibinfo {author} {\bibfnamefont {A.~D.}\ \bibnamefont {C{\'o}rcoles}},
  \bibinfo {author} {\bibfnamefont {A.}~\bibnamefont {Mezzacapo}}, \bibinfo
  {author} {\bibfnamefont {J.~M.}\ \bibnamefont {Chow}},\ and\ \bibinfo
  {author} {\bibfnamefont {J.~M.}\ \bibnamefont {Gambetta}},\ }\bibfield
  {title} {\bibinfo {title} {Error mitigation extends the computational reach
  of a noisy quantum processor},\ }\href
  {https://doi.org/10.1038/s41586-019-1040-7} {\bibfield  {journal} {\bibinfo
  {journal} {Nature}\ }\textbf {\bibinfo {volume} {567}},\ \bibinfo {pages}
  {491} (\bibinfo {year} {2019})}\BibitemShut {NoStop}%
\end{thebibliography}%

\onecolumngrid
\appendix
\section{Review of quantum error mitigation methods}
In this section, we briefly review the quantum error mitigation methods used in this paper: polynomial extrapolation method and virtual distillation.
First, we review the error extrapolation method for physical errors.
Then, we explain purification-based quantum error mitigation methods, which use multiple copies of noisy quantum states.

\subsection{Error-extrapolation methods for physical errors}
We review the polynomial error extrapolation according to Refs.~\cite{liEfficientVariationalQuantum2017,temmeErrorMitigationShortDepth2017,endoHybridQuantumClassicalAlgorithms2021,caiQuantumErrorMitigation2021}.
Suppose that we measure the expectation value of an observable $A$.
We assume that there is one parameter for the physical error $p$ in a quantum circuit.
Expanding an expectation value of an observable $\ev{A}\pqty{p}$ up to the $L$th order, we obtain
\begin{align}
    \ev{A}\pqty{p}
    =& \ev{A}\pqty{0} + \sum_{i=1}^L A_{i}p^i + \mathcal{O}\pqty{p^{L+1}} \label{eq:expansion_of_ev_1D_physical_error_extrapolation},
\end{align}
where $A_i$ stand for the coefficients of the expansion.
Considering the $L+1$ expectation values $\ev{A}\pqty{p_1}, \ldots, \ev{A}\pqty{p_{L}}$ and expanding them in the same way as in \cref{eq:expansion_of_ev_1D_physical_error_extrapolation},
we have the following estimator for the polynomial extrapolation:
\begin{align}
    \ev{A}_\text{est}\pqty{0} = \sum_{i=0}^L\ev{A}\pqty{p_i}\prod_{j \neq i} \frac{p_j}{p_j-p_i}.
\end{align}
The variance of this estimator is given by
\begin{align}
    \text{Var}\bqty{\ev{A}_\text{est}\pqty{0}} = \sum_{i=0}^L\text{Var}\bqty{\ev{A}\pqty{p_i}}\pqty{\prod_{j \neq i} \frac{p_j}{p_j-p_i}}^2.
\end{align}

\subsubsection{Exponential error extrapolation}
It is pointed out that the exponential error extrapolation is more suitable than the polynomial error extrapolation if the number of quantum gates $N_g$ is large and the error rate of quantum noise $p$ is small.
Suppose that all of the quantum gates suffer from the same quantum noise defined as
\begin{align}
    \mathcal{N} = \pqty{1-p}\bqty{I} + p \mathcal{D},
\end{align}
where $\bqty{A}\pqty{\rho} \coloneqq A \rho A^\dag$ and $\mathcal{D}$ stands for a noisy map.
Then, the noisy quantum circuit can be represented as
\begin{align}
    \mathcal{E}_\text{circ} 
    &= \prod_{i=1}^{N_g} \mathcal{N}_i \circ \mathcal{U}_i\\
    &= \sum_{i=0}^{N_g} \pqty{1-p}^{N_g-i} p^i \sum_{j=1}^{\binom{N_g}{i}}\mathcal{G}_i^j, \label{eq:noisy_circ_decomposition}
\end{align}
where $\mathcal{U}_i$ stands for an ideal quantum gate and $\mathcal{G}_i^j$ stands for one of the expansion with $i$ errors.
Denoting the average of $\mathcal{G}_i^j$ as 
\begin{align}
    \mathcal{G}_i = \frac{\sum_{j=1}^{\binom{N_g}{i}}\mathcal{G}_i^j}{\binom{N_g}{i}},
\end{align}
\cref{eq:noisy_circ_decomposition} can be rewritten as
\begin{align}
    \mathcal{E}_\text{circ} = \sum_{i=0}^{N_g} \pqty{1-p}^{N_g-i}p^i\binom{N_g}{i}\mathcal{G}_i.\label{eq:noisy_circ_binomial}
\end{align}
Since $\pqty{1-p}^{N_g-i}p^i\binom{N_g}{i}$ is the binomial distribution, if $N_g$ is sufficiently large and $p$ is sufficiently small satisfying $N_gp = \mathcal{O}\pqty{1}$, the binomial distribution can be approximated by the Poisson distribution $e^{-N_gp}\frac{\pqty{N_gp}^i}{i!}$.
Thus, \cref{eq:noisy_circ_binomial} can also be approximated by
\begin{align}
    \mathcal{E}_\text{circ} \approx e^{-N_gp}\sum_{i=0}^{N_g} \frac{\pqty{N_gp}^i}{i!}\mathcal{G}_i.\label{eq:noisy_circ_poisson}
\end{align}
From \cref{eq:noisy_circ_poisson}, the noisy expectation values $\ev{A}\pqty{p}$ and $\ev{A}\pqty{rp}$ $\pqty{r>1}$ can be given by
\begin{align}
    \ev{A}\pqty{p} &= e^{-p} \ev{A}\pqty{0},\\
    \ev{A}\pqty{rp} &= e^{-rp} \ev{A}\pqty{0}.
\end{align}
Thus, we have the following estimator for exponential extrapolation:
\begin{align}
    \ev{A}_\text{est}\pqty{0} = \ev{A}\pqty{p}^{\frac{r}{r-1}} \ev{A}\pqty{rp}^{\frac{1}{1-p}}
\end{align}

\subsection{Purification-based quantum error mitigation}
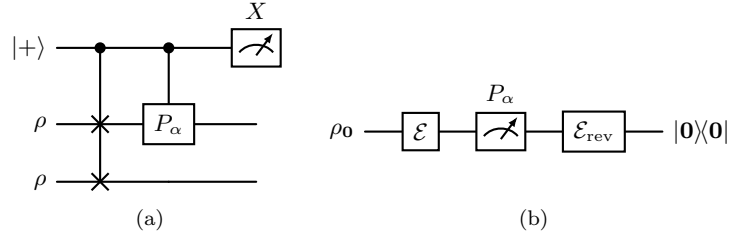
\begin{figure}[tb]
\centering
\subfloat[\label{subfig:copy_based_purification}]{%
    \begin{adjustbox}{valign=b}
        \begin{quantikz}
            \lstick{$\ket{+}$} & \ctrl{1} & \ctrl{1} & \meter{X}\\
            \lstick{$\rho$} & \swap{1} & \gate{P_\alpha} &\\
            \lstick{$\rho$} & \targX{} & &
        \end{quantikz}
    \end{adjustbox}
    } \quad
\subfloat[\label{subfig:dual_state_purification}]{%
    \begin{adjustbox}{valign=b}
        \begin{quantikz}
            \lstick{$\rho_{\bm{0}}$} & \gate{\mathcal{E}} & \meter{P_\alpha} & \gate{\mathcal{E}_\text{rev}} & \rstick{$\ketbra{\bm{0}}{\bm{0}}$}\\
        \end{quantikz}
    \end{adjustbox}
    }
        \caption{
        Quantum circuits for purification-based QEM methods.
        (a): Copy-based purification.
        $\text{Tr}\pqty{P_\alpha\rho^2}$ can be obtained by the measurement of Pauli $X$ of the most upper wire.
        (b): Dual-state purification with mid-circuit measurement.
        $\text{Tr}\pqty{\frac{\bar{\rho}\rho + \rho \bar{\rho}}{2}P_\alpha}$ can be obtained by mid-circuit measurement of Pauli operator $P_\alpha$ and then postselecting on the measurement outcome $\bm{0}$, where $\bar{\rho}$ is the dual state of $\rho$.
        $\text{Tr}\pqty{\frac{\bar{\rho}\rho + \rho \bar{\rho}}{2}P}$ is expected to be a good approximation of $\text{Tr}\pqty{P\rho^2}$.
        }
    \label{fig:various_purification_based_QEM}
\end{figure}
Purification-based QEM methods use several copies of noisy state $\rho$ to obtain an expectation value of an observable for a "purified" state, and it is suited for mitigating stochastic errors \cite{hugginsVirtualDistillationQuantum2021,koczorExponentialErrorSuppression2021,koczorDominantEigenvectorNoisy2021}.
This method simulates the purified state to estimate an expectation value at the cost of state copies or the circuit depth.
Let $A$ be an observable.
Then, the estimator for the expectation value of $A$ when using the purification-based QEM is given by
\begin{align}
    \ev{A}_\text{est} = \frac{\text{Tr}\pqty{\rho^L A}}{\text{Tr}\pqty{\rho^L}} \label{eq:estimator_of_VD}.
\end{align}
The corresponding effective state could be written as
\begin{align}
    \rho_\text{PB} = \frac{\rho^L}{\text{Tr}\pqty{\rho^L}},\label{eq:effective_state_of_VD}
\end{align}
for the integer $L\geq 2$.
Now, let the spectral decomposition of the noisy state denote $\rho=\sum_k p_k \ket{\phi_k}\bra{\phi_k}\quad \pqty{p_0 \geq p_1 \geq \ldots}$, $\langle \phi_i | \phi_j \rangle = \delta_{ij}$.
Then, we get
\begin{align}
    \rho^L= p_0^L \Bqty{\ket{\phi_0}\bra{\phi_0} +\sum_{k \geq 1} \pqty{\frac{p_k}{p_0}}^L \ket{\phi_k}\bra{\phi_k}},
\end{align}
which clearly shows the exponential suppression of the effect of subdominant eigenvectors.
Because it has been shown that the dominant eigenvector well approximates the noiseless state in the presence of stochastic errors such as amplitude damping and stochastic Pauli errors,
this method can mitigate stochastic errors efficiently \cite{koczorDominantEigenvectorNoisy2021,hugginsVirtualDistillationQuantum2021}.

There are two types of implementations for the purification-based QEM: copy-based purification \cite{hugginsVirtualDistillationQuantum2021,koczorExponentialErrorSuppression2021,koczorDominantEigenvectorNoisy2021} and dual-state purification \cite{huoDualstatePurificationPractical2022,caiResourceefficientPurificationbasedQuantum2021,yangResourceefficientGeneralizedQuantum2025}.
In the former purification protocol, we use copies of the quantum state.
To measure $\text{Tr}\pqty{\rho^L P_\alpha}$ for a Pauli operator $P_\alpha$ for $L=2$,
we resort to the Hadamard test circuit in which the target unitary is the product of the swap operator and the observable of interest, as shown in \cref{subfig:copy_based_purification}.
Also, we can compute the denominator by using the circuit that removes the controlled-$P_\alpha$ operator from the circuit shown in \cref{subfig:copy_based_purification}.
See Refs.~\cite{hugginsVirtualDistillationQuantum2021,koczorExponentialErrorSuppression2021,koczorDominantEigenvectorNoisy2021} for details.

Meanwhile, the dual-state purification method purifies the error-mitigated state with the doubled-depth quantum circuit as shown in \cref{subfig:dual_state_purification}.
Let the initial state denote $\rho_{\bm{0}} = \ketbra{\bm{0}}{\bm{0}}$, and let the process for the computation denote $\mathcal{E}$ that turns $\rho_{\bm{0}}$ into $\rho = \mathcal{E}\pqty{\rho_{\bm{0}}}$.
We also introduce the uncomputational process $\mathcal{E}_\text{rev}$ corresponding to $\mathcal{E}$.
More precisely, $\mathcal{E}$ and $\mathcal{E}_\text{rev}$ are described as
\begin{align}
    \mathcal{E}&\coloneqq \mathcal{N}_L \circ \mathcal{U}_L \circ \cdots \circ \mathcal{N}_1 \circ \mathcal{U}_1, \\
    \mathcal{E}_\text{rev}&\coloneqq \mathcal{N}_1^{\pqty{\text{rev}}} \circ \mathcal{U}_1 \circ \cdots \circ \mathcal{N}_L^{\pqty{\text{rev}}} \circ \mathcal{U}_L,
\end{align}
where $\mathcal{U}_k$ denotes an error-free quantum gate process, and $\mathcal{N}_k$ or $\mathcal{N}_k^{\pqty{\text{rev}}}$ denotes the accompanying quantum noise process.
Then, the estimator for the expectation value of $A$ when using the dual-state purification is given by
\begin{align}
    \ev{A}_\text{est} = \frac{\text{Tr}\pqty{\frac{\rho\bar{\rho}+\bar{\rho}\rho}{2}A}}{\text{Tr}\pqty{\rho\bar{\rho}}}\label{eq:estimator_of_dual_state_purification}.
\end{align}
Here, $\bar{\rho} \coloneqq \mathcal{E}_\text{rev}^\dag (\rho_{\bf{0}})=\sum_i K_i^\dag \rho_{\bf{0}} K_i $ is called the dual state, with $\mathcal{E}_\text{rev}\pqty{\sigma}=\sum_i K_i \sigma K_i^\dag$.
In the case without noise, we have $ \rho= \bar{\rho}$, and thus it is highly likely the dual state is a good approximation of the state $\rho$;
furthermore, assuming that $\mathcal{N}_k = \mathcal{N}_k^{\pqty{\text{rev}}}$ for all $k$, it has been numerically shown that the trace distance of $\rho$ and $\bar{\rho}$ becomes small under stochastic Pauli noise as the depth of the quantum circuit increases \cite{yangResourceefficientGeneralizedQuantum2025}.

We evaluate the joint probability of measuring the $x \in \Bqty{0, 1 }$ in the mid-circuit measurement and measuring all the qubits in $0$ at the final measurement.
The joint probability reads $p_{x, \bm{0}}=\text{Tr}\bqty{\mathcal{E}_\text{rev}\pqty{\frac{I+ (-1)^x P_a }{2} \mathcal{E}(\rho_{\bm{0}}) \frac{I+ (-1)^x P_a }{2} } P_{\bm{0}} }$ for the initial state $\rho_{\bm{0}}=\ket{\bm{0}}\bra{\bm{0}}$ and the projector onto the initial state $P_{\bm{0}}=\ket{\bm{0}}\bra{\bm{0}}$.
Then, we can show
\begin{align}
    p_{0, \bf{0}}- p_{1, \bf{0}} &= \text{Tr}\pqty{\frac{\bar{\rho}\rho+\rho 
    \bar{\rho}}{2} P_a}.\label{eq:dual_state_purification_mid} 
\end{align}
Thus, we can evaluate the numerator.
Also, by removing the mid-circuit measurement from the circuit shown in \cref{subfig:dual_state_purification}, we can evaluate the denominator by estimating the probability of measuring all the qubits in 0.
Indeed, the probability is given by
\begin{align}
    p_{\bm{0}} = \text{Tr}\bqty{\mathcal{E}_\text{rev}\pqty{\mathcal{E}(\rho_{\bm{0}})} P_{\bm{0}} } = \text{Tr}\pqty{\rho\bar{\rho}}.
\end{align}

\section{General remarks of error-extrapolation methods}\label{sec:general_remark_of_ext}
Error extrapolation methods are relatively easier to implement than the other QEM methods, and thus, various experiments have been conducted \cite{kandalaErrorMitigationExtends2019,kimEvidenceUtilityQuantum2023}.
However, the effective state of an error extrapolation method may not be physical in general, which may cause bias. 
For example, consider the following estimator of extrapolation methods: 
\begin{align}
    \ev{A}_\text{est} 
    &= \sum_i c_i \ev{A}\pqty{p_i}\\
    &= \sum_i c_i \text{Tr}\pqty{\rho\pqty{p_i} A} .\label{eq:estimator_of_error_extrapolation}
\end{align}
The error extrapolation methods, such as the polynomial extrapolation and the exponential extrapolation, can be written in the above form.
From \cref{eq:estimator_of_error_extrapolation}, the corresponding effective state could be written as
\begin{align}
    \rho_\text{extra} = \sum_i c_i \rho_i. \label{eq:effective_state_of_extrapolation}
\end{align}
In general, $\rho_\text{extra}$ could not satisfy $\rho\geq 0$.

An estimated expectation value using an unphysical state may cause a large bias.
Consider an expectation value of an observable $A$ for a general operator $\rho$.
The expectation value can be bounded by the following inequality:
\begin{align}
    \abs{\text{Tr}\pqty{A\rho}}\leq \abs{\rho}_\text{op}\text{Tr}\pqty{\abs{A}} \label{eq:bound_of_ev_for_general_op},
\end{align}
where $\abs{\cdot}_\text{op}$ is the operator norm, i.e., the maximum singular value of the operator $A$.
If $\rho$ is a density operator, $\abs{\rho}_\text{op}\leq 1$, and thus we obtain
\begin{align}
    \abs{\text{Tr}\pqty{A\rho}}\leq \text{Tr}\pqty{\abs{A}}\label{eq:bound_of_ev_for_density_op}.
\end{align}
From \cref{eq:bound_of_ev_for_density_op}, we see that the absolute value of an estimated expectation value for a physical state is upper-bounded by the trace norm of $A$ and independent of $\rho$.
However, from \cref{eq:bound_of_ev_for_general_op}, if $\rho$ is an unphysical state and has a large negative eigenvalue, the absolute value could be larger than the trace norm of $A$.

\section{Quantum circuits for Trotter subspace expansion}\label{sec:explanation_of_circuits}
In this section, we explain how the expectation value of an observable $A=\sum_\alpha c_\alpha P_\alpha$ for the ansatz state of the Trotter subspace expansion can be evaluated in detail.
First, we show the evaluation method using the standard purification method, i.e., virtual distillation.
Then, we explain the evaluation methods using dual-state purification methods, which require half the quantum states of the standard purification method at the expense of the circuit depth.

\subsection{Standard purification method}
In the case of the standard purification method, we adopt the ansatz state shown in \cref{eq:Trotter_ansatz}.
Recall that the expectation value of $A$ for the ansatz state can be given by
\begin{align}
    \ev{A}_\text{est} = 
    \frac{\sum_{i,j=0}^{n'} g_i g_j \text{Tr}\pqty{\rho_i\rho_j A}}{\sum_{i,j=0}^{n'} g_i g_j \text{Tr}\pqty{\rho_i\rho_j}}
    =
    \frac{\sum_\alpha c_\alpha \sum_{i,j=0}^{n'} g_i g_j \text{Tr}\pqty{\rho_i\rho_j P_\alpha}}{\sum_{i,j=0}^{n'} g_i g_j \text{Tr}\pqty{\rho_i\rho_j}},
\end{align}
we have to evaluate the numerator and the denominator.
To evaluate these expectation values, we review how the quantum circuits in \cref{fig:qc_for_evaluating_of_ev_of_observarble} can measure $\text{Tr}\pqty{\rho_i\rho_j}$ and $\text{Tr}\pqty{\rho_i\rho_jP_\alpha}$, where $\rho_i \coloneqq \rho\pqty{p_i,M_i}$.

The expectation value of the Pauli $X$ measurement in the circuit shown in \cref{subfig:swap_test} $\ev{X\otimes I^{\otimes 2}}$ is given by
\begin{align}
    \ev{X\otimes I^{\otimes 2}}_I
    &= \frac{1}{2}\text{Tr}\pqty{X \otimes I^{\otimes 2} \ketbra{0}{1} \otimes \text{SWAP}\pqty{\rho_i \otimes \rho_j} + \text{h.c.}}\\
    &= \text{Tr}\pqty{\rho_i\rho_j}\label{eq:X_measure_for_fidelity}.
\end{align}
From \cref{eq:X_measure_for_fidelity}, we find that the denominator can be estimated by the circuit shown in \cref{subfig:swap_test}.

The expectation value of the Pauli $X$ measurement in the circuit shown in \cref{subfig:circ_for_hij} $\ev{X\otimes I^{\otimes 2}}$ is given by
\begin{align}
    \ev{X\otimes I^{\otimes 2}}_{P_\alpha}
    &= \frac{1}{2}\text{Tr}\pqty{X \otimes P_\alpha \otimes I \ketbra{0}{1} \otimes \text{SWAP}\pqty{\rho_i \otimes \rho_j} + \text{h.c.}}\\
    &= \text{Tr}\pqty{\frac{\rho_i\rho_j + \rho_j\rho_i}{2}P_\alpha}\label{eq:X_measure_for_exp_of_obs}.
\end{align}
On the other hand, the numerator of $\ev{A}_\text{est}$ can be rewritten as
\begin{align}
    \sum_\alpha c_\alpha \sum_{i,j=0}^{n'} g_i g_j \text{Tr}\pqty{\rho_i\rho_j P_\alpha}
    &=\sum_\alpha c_\alpha \bqty{\sum_{i=0}^{n'} g_i^2 \text{Tr}\pqty{\rho_i^2 P_\alpha} + \sum_{i<j} g_ig_j\Bqty{\text{Tr}\pqty{\rho_i\rho_j P_\alpha} + \text{Tr}\pqty{\rho_j\rho_i P_\alpha}}}\\
    &=\sum_\alpha c_\alpha \Bqty{\sum_{i=0}^{n'} g_i^2 \text{Tr}\pqty{\rho_i^2 P_\alpha} + 2\sum_{i<j} g_ig_j\text{Tr}\pqty{\frac{\rho_i\rho_j + \rho_j\rho_i}{2}P_\alpha}}. \label{eq:numerator_in_the_form_of_X_measurement}
\end{align}
Combining \cref{eq:X_measure_for_exp_of_obs} with \cref{eq:numerator_in_the_form_of_X_measurement},
we find that both $\text{Tr}\pqty{\rho_i^2P_\alpha}$ and $\text{Tr}\pqty{\frac{\rho_i\rho_j + \rho_j \rho_i}{2}P_\alpha}$ can be estimated by the circuit shown in \cref{subfig:circ_for_hij}, and thus the numerator can also be estimated by the circuit.

\subsection{Dual-state purification method}
In the case of the dual-state purification method, there are two choices of the ansatz states: $\rho_\text{Dual}^{\pqty{1}}$ and $\rho_\text{Dual}^{\pqty{2}}$ defined as \cref{eq:dual1,eq:dual2}.
After reviewing the measurement outcomes of the quantum circuits in \cref{fig:dual_qc_for_evaluating_of_ev_of_observarble},
we explain how we can evaluate the expectation value of an observable $A=\sum_\alpha c_\alpha P_\alpha$ for such an ansatz state.

\Cref{subfig:dual_se_intermediate} shows the quantum circuit for the Trotter subspace expansion using the dual-state purification with the mid-circuit measurement.
In this circuit, we execute the mid-circuit measurement and then post-select the output state on the measurement outcome $\bm{0}$.
Let $p_{x,\bm{0}}$ be the joint probability of measuring $x \in \Bqty{0,1}$ in the mid-circuit measurement.
The joint probability $p_{x, \bm{0}}$ is
\begin{align}
    p_{x, \bm{0}}
    &= \text{Tr}\bqty{\mathcal{E}_{\text{rev}\ j}\pqty{\frac{I+ (-1)^x P_a }{2} \mathcal{E}_i\pqty{\rho_{\bm{0}}} \frac{I+ (-1)^x P_a }{2} } P_{\bm{0}} }\\
    &= \text{Tr}\bqty{\frac{I+ (-1)^x P_a }{2} \rho_i \frac{I+ (-1)^x P_a }{2} \bar{\rho}_j }, \label{eq:joint_prob_dual_mid}
\end{align}
where $\mathcal{E}_{\text{rev}\ j}\coloneqq \mathcal{E}_\text{rev}\pqty{p_j,M_j}$, $\mathcal{E}_i\coloneqq\mathcal{E}\pqty{p_i, M_i}$, $\rho_{\bm{0}}\coloneqq\ketbra{\bm{0}}{\bm{0}}$, and $P_{\bm{0}}\coloneqq \ketbra{\bm{0}}{\bm{0}}$.
From \cref{eq:joint_prob_dual_mid}, we obtain the following equation:
\begin{align}
    p_{0,\bm{0}} - p_{1,\bm{0}} &= \text{Tr}\pqty{\frac{\rho_i\bar{\rho}_j + \bar{\rho}_j\rho_i}{2}P_\alpha} \label{eq:expectation_value_of_mid_circuit_measurement}.
\end{align}
Also, by removing the mid-circuit measurement from the circuit shown in \cref{subfig:dual_se_intermediate}, the probability of measuring all the qubits in 0 is given by
\begin{align}
    p_{\bm{0}} = \text{Tr}\pqty{\rho_i\bar{\rho}_j}. \label{eq:probability_of_all_the_qubits_0_using_mid_circuit_measurement}
\end{align}

\Cref{subfig:dual_se_indirect} shows the quantum circuit for the Trotter subspace expansion using the dual-state purification with the indirect measurement.
The measurement outcomes become
\begin{align}
    \ev{X\otimes P_{\bm{0}}} &= \frac{1}{2}\Bqty{\text{Tr}\pqty{\mathcal{E}_{\text{rev}\ j}\pqty{\mathcal{E}_i\pqty{\rho} P_\alpha}\ketbra{\bm{0}}{\bm{0}}}
    + \text{Tr}\pqty{\mathcal{E}_{\text{rev}\ j}\pqty{P_\alpha \mathcal{E}_i\pqty{\rho}}\ketbra{\bm{0}}{\bm{0}}}} 
    = \text{Tr}\pqty{\frac{\rho_i\bar{\rho}_j + \bar{\rho}_j\rho_i}{2}P_\alpha}, \label{eq:expectation_value_of_indirect_X_measurement}\\
    \ev{Y\otimes P_{\bm{0}}} &= \frac{1}{2}\Bqty{-i\text{Tr}\pqty{\mathcal{E}_{\text{rev}\ j}\pqty{\mathcal{E}_i\pqty{\rho} P_\alpha}\ketbra{\bm{0}}{\bm{0}}}
    + i\text{Tr}\pqty{\mathcal{E}_{\text{rev}\ j}\pqty{P_\alpha \mathcal{E}_i\pqty{\rho}}\ketbra{\bm{0}}{\bm{0}}}} 
    = i\text{Tr}\pqty{\frac{\rho_i\bar{\rho}_j - \bar{\rho}_j\rho_i}{2}P_\alpha}.\label{eq:expectation_value_of_indirect_Y_measurement}
\end{align}
Combining these results, we see that
\begin{align}
    \ev{X\otimes P_{\bm{0}}} - i \ev{Y\otimes P_{\bm{0}}} = \text{Tr}\pqty{\rho_i \bar{\rho}_j P_\alpha}.\label{eq:estimator_of_numerator_using_dual_two}
\end{align}

\subsubsection{\texorpdfstring{$\rho_\text{Dual}^{\pqty{1}}$}{Dual 1 ansatz state}}
The expectation value of an observable for $\rho_\text{Dual}^{\pqty{1}}$ is given by
\begin{align}
    \ev{A}_\text{est} = \frac{\sum_\alpha c_\alpha \sum_{i,j}g_ig_j\text{Tr}\pqty{\frac{\rho_i\bar{\rho}_j+\bar{\rho}_j\rho_i}{2}P_\alpha}}{\sum_{i,j}\text{Tr}\pqty{\rho_i \bar{\rho}_j}}. \label{eq:estimator_of_TSE_using_dual_one}
\end{align}
According to \cref{eq:expectation_value_of_mid_circuit_measurement,eq:expectation_value_of_indirect_X_measurement}, the numerator can be estimated by the circuit using the mid-circuit measurement shown in \cref{subfig:dual_se_intermediate} or the circuit using the indirect measurement shown in \cref{subfig:dual_se_indirect}.
Also, from \cref{eq:probability_of_all_the_qubits_0_using_mid_circuit_measurement}, the denominator can be estimated by the circuit that removes the mid-circuit measurement from the one shown in \cref{subfig:dual_se_intermediate}.

\subsubsection{\texorpdfstring{$\rho_\text{Dual}^{\pqty{2}}$}{Dual 2 ansatz state}}
The expectation value of an observable for $\rho_\text{Dual}^{\pqty{2}}$ is given by
\begin{align}
    \ev{A}_\text{est} = \frac{\sum_\alpha c_\alpha \sum_{i,j}g_ig_j\text{Tr}\pqty{\rho_i\bar{\rho}_j P_\alpha}}{\sum_{i,j}\text{Tr}\pqty{\rho_i \bar{\rho}_j}}. \label{eq:estimator_of_TSE_using_dual_two}
\end{align}
According to \cref{eq:estimator_of_numerator_using_dual_two}, the numerator can be estimated by the circuit using the indirect measurement shown in \cref{subfig:dual_se_indirect}.
Note that we need to perform measurements of not only $X$ but also $Y$ on the ancilla qubit in this circuit, contrary to the case of $\rho_\text{Dual}^{\pqty{1}}$, where we only need to perform a measurement of $X$.
Similar to the case of $\rho_\text{Dual}^{\pqty{1}}$, the denominator can be estimated by the circuit that removes the mid-circuit measurement from the one shown in \cref{subfig:dual_se_intermediate}.

\section{Validation of \cref{eq:opt_trotter_1d_tfi} by numerical simulation}\label{sec:scaling_of_coeff_of_opt_Trotter_num}
\begin{figure}
    \centering
     \subfloat[
            \label{subfig:scaling_of_opt_Trotter_num}]{%
        \includegraphics[width=0.48\columnwidth]{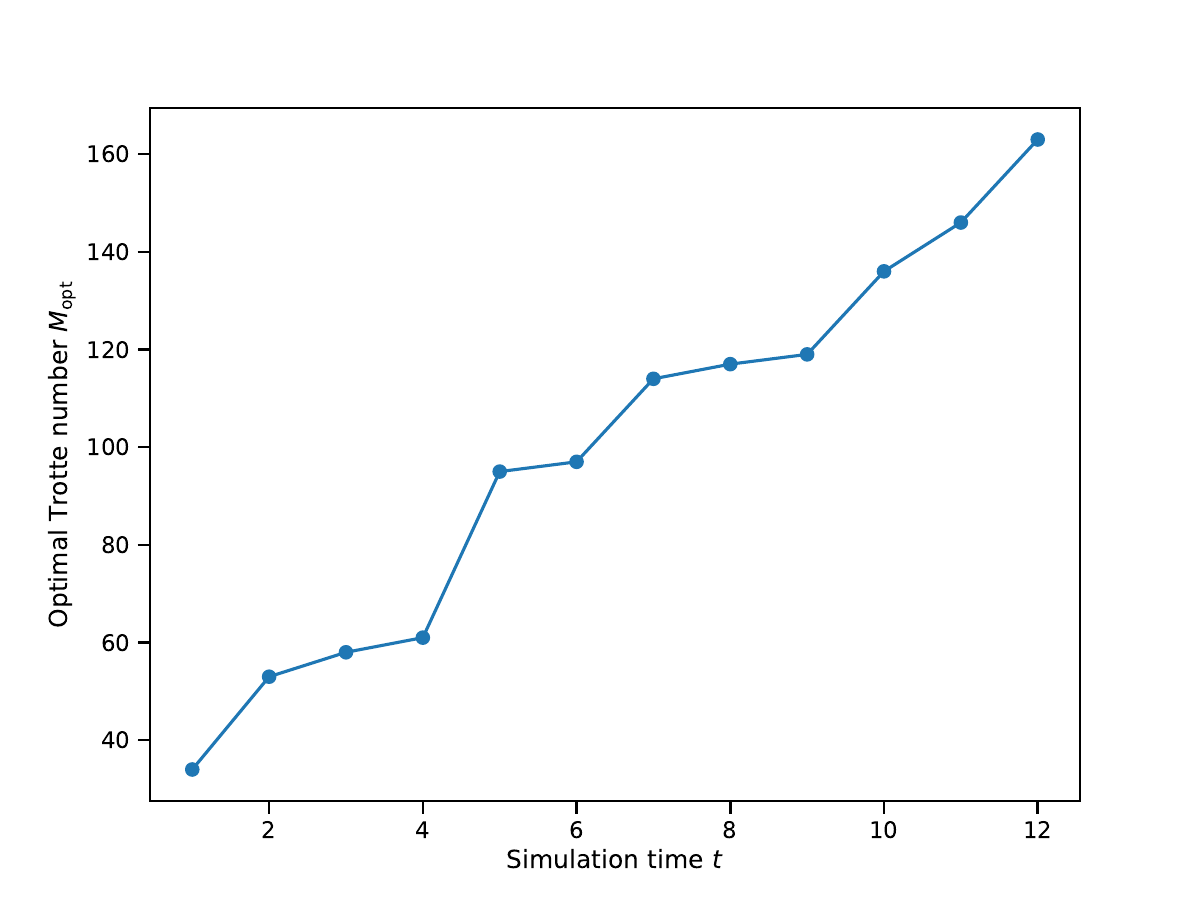}%
    }
    \hfill
     \subfloat[
     \label{subfig:scaling_of_coefficient_of_opt_Trotter_num}]{%
        \includegraphics[width=0.48\columnwidth]{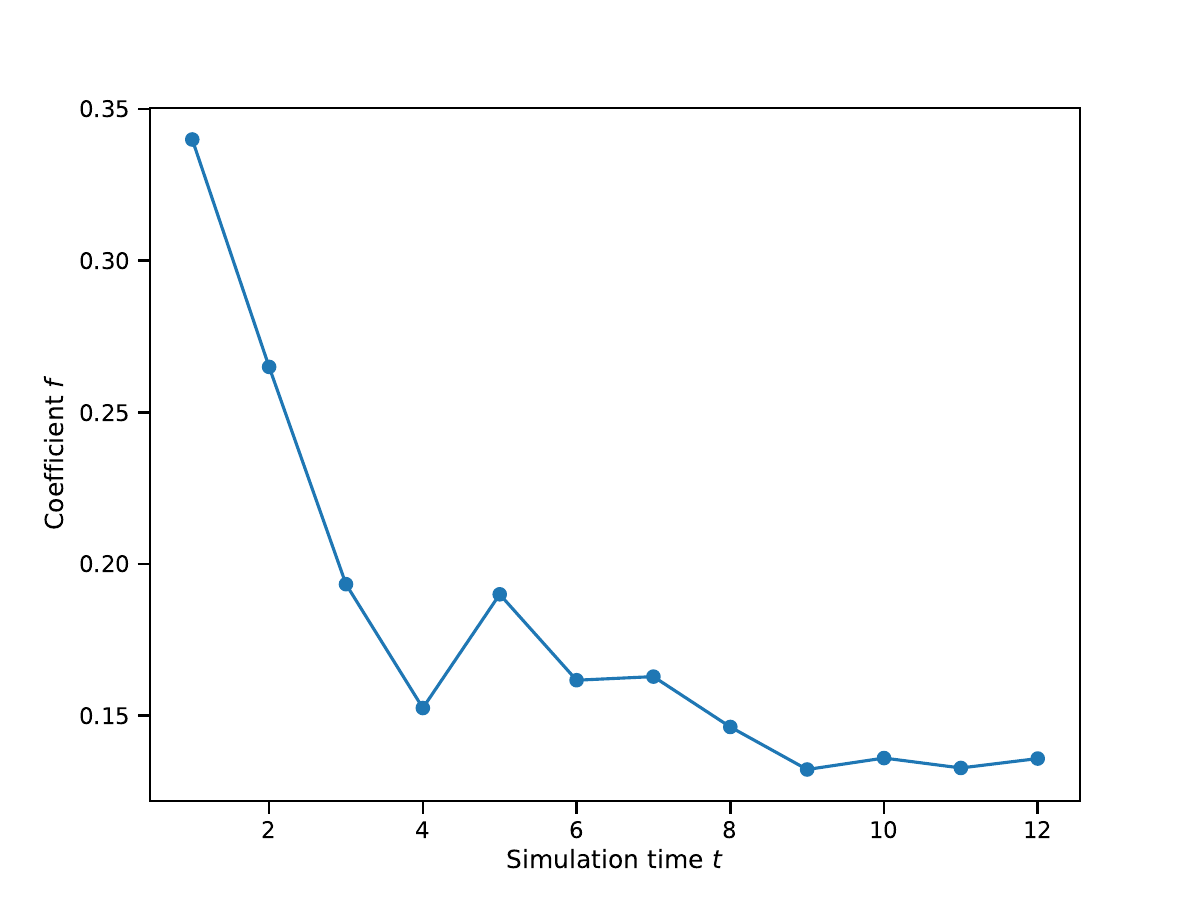}%
    }
    \caption{
    Scaling of (a) the optimal Trotter number $M_\text{opt}$ and (b) its coefficient $f$ as a function of the simulation time $t$.
    In our numerical calculation, we set $n = 6$, $p_1 = 1 \times 10^{-5}$, $p_2 = 1 \times 10^{-4}$, $J = -1$, $B = 1$, and $t = 1, 2, ..., 12$.
    }
\end{figure}%
Here, we analyze the scaling of the coefficient of the optimal Trotter number $M_\text{opt}$ as a function of the simulation time $t$ to validate \cref{eq:opt_trotter_1d_tfi}.
In \cref{subfig:scaling_of_opt_Trotter_num}, we show that the scaling of the optimal Trotter step $M_\text{opt}$ for the dynamics of 1D TFI as a function of the simulation time $t$, where $M_\text{opt}$ is obtained numerically.
It is observed that the optimal Trotter step increases as the simulation time increases, as expected.
Then, we also calculate the scaling of the coefficient of the optimal Trotter number $f$ as a function of the simulation time $t$, where $f$ can be given by
\begin{align}
    f = \frac{M_\text{opt}}{\abs{t}\sqrt{\frac{\abs{J}\abs{B}}{p_2}}}.
\end{align}
From \cref{subfig:scaling_of_coefficient_of_opt_Trotter_num}, we confirm that $f$ decreases until $t=8$, but remains constant thereafter.
Thus, our derived scaling of the optimal Trotter number shown in \cref{eq:opt_trotter_1d_tfi} can be valid when the simulation time is sufficiently long.

As discussed above, the coefficient for the optimal Trotter number can be obtained by combining theoretical analysis with numerical calculations.
We also expect that regression analysis would be needed when we execute quantum simulations that are intractable for classical computers using actual quantum computers.

\section{Shot noise on observable estimation by Trotter subspace expansion}\label{sec:shot_noise}
In the previous section, we have explained that the quantum circuits in \cref{fig:qc_for_evaluating_of_ev_of_observarble,fig:dual_qc_for_evaluating_of_ev_of_observarble} with an infinite number of measurements give the expectation values for the ansatz states.
In practice, expectation values of observables differ from the ideal ones because the total number of measurements is finite, which is called the shot noise.
The expectation value of an observable $A$ with finite measurements could be expressed as $\ev{A}_\text{est} + \delta A_\text{est}$, where $\delta A_\text{est}$ stands for the deviation due to the shot noise.
In our numerical simulation, we consider $\delta A_\text{est}$ by adding the Gaussian noise whose amplitude is determined by $\text{Var}\bqty{A_\text{est}}$
to the expectation value without shot-noise.
$\text{Var}\bqty{A_\text{est}}$ is composed of each variance of the Pauli observables shown in \cref{fig:qc_for_evaluating_of_ev_of_observarble}.
More precisely, in our numerical simulations, we consider the shot noise according to the following procedure:
\begin{enumerate}
    \item Compute an expectation value of all of the Pauli observables shown in \cref{fig:qc_for_evaluating_of_ev_of_observarble} or \cref{fig:dual_qc_for_evaluating_of_ev_of_observarble} without shot noise.
    \item Compute the corresponding single-shot variances: 
    $\text{Var}\bqty{\pqty{X\otimes I^{\otimes 2}}_{P_\alpha}}$, $\text{Var}\bqty{\pqty{Y\otimes I^{\otimes 2}}_{P_\alpha}}$, and $\text{Var}\bqty{\pqty{X\otimes I^{\otimes 2}}_{I}}.$\label{item:pauli_shot_noise}
    \item Generate the Gaussian noise $\mathcal{N}\pqty{0,V_\text{single}/N_\text{circ}}$, where $V_\text{single}$ and $N_\text{circ}$ denote one of the single-shot variances in \cref{item:pauli_shot_noise} and the number of measurements allocated for one quantum circuit to obtain one expectation value, respectively.
    \item Add the Gaussian noise to the ideal expectation value.\label{item:adding_shot_noise}
    \item Using the results from \cref{item:adding_shot_noise}, output the estimator.
\end{enumerate}
In the following, we will show the single-shot variance of the Pauli observables and then theoretically analyze the single-shot variance of the estimator with the Trotter subspace expansion.

\subsection{Single-shot variance of Pauli observables}
We discuss the single-shot variance of Pauli observables $P_\alpha$ or the identity $I$ using the quantum circuits shown in \cref{fig:qc_for_evaluating_of_ev_of_observarble}: $\text{Var}\bqty{\pqty{X\otimes I^{\otimes 2}}_{P_\alpha}}$, $\text{Var}\bqty{\pqty{Y\otimes I^{\otimes 2}}_{P_\alpha}}$, and $\text{Var}\bqty{\pqty{X\otimes I^{\otimes 2}}_{I}}$.
By definition, we obtain
\begin{align}
    \text{Var}\bqty{\pqty{X\otimes I^{\otimes 2}}_{P_\alpha}}
    &= \ev{\pqty{X\otimes I^{\otimes 2}}^2_{P_\alpha}} - \ev{X\otimes I^{\otimes 2}}^2_{P_\alpha}\\
    &= \ev{I^{\otimes 3}}_{P_\alpha} - \ev{X\otimes I^{\otimes 2}}^2_{P_\alpha}\\
    &= 1 - \ev{X\otimes I^{\otimes 2}}^2_{P_\alpha}. \label{eq:single_shot_variance_of_pauli}
\end{align}

Also, the single-shot variance $\text{Var}\bqty{\pqty{X\otimes I^{\otimes 2}}_{I}}$ is given by
\begin{align}
    \text{Var}\bqty{\pqty{X\otimes I^{\otimes 2}}_{I}}
    &= \ev{I^{\otimes 3}}_I - \ev{X\otimes I^{\otimes 2}}^2_{I} \\
    &= 1 - \text{Tr}\pqty{\rho_i\rho_j}^2,\label{eq:single_shot_variance_of_fidelity}
\end{align}
where we use \cref{eq:X_measure_for_fidelity}.

\subsection{Single-shot variance of the estimator with the Trotter subspace expansion}\label{sec:single_shot_variance_of_tse}
Here we give the theoretical analysis of the single-shot variance of the estimator with the Trotter subspace expansion.
The expectation value for the ansatz state using the Trotter subspace is given by
\begin{align}
    \ev{A}_\text{est} = \frac{\sum_\alpha c_\alpha \pqty{\sum_{i=0}^{n'} g_i^2 \text{Tr}\pqty{\rho_i^2 P_\alpha} + 2\sum_{i<j} g_ig_j\text{Tr}\pqty{\frac{\rho_i\rho_j + \rho_j\rho_i}{2}P_\alpha}}}{\sum_{i,j=0}^{n'} g_i g_j \text{Tr}\pqty{\rho_i\rho_j}}.
\end{align}
Using the quantum circuits shown in \cref{fig:qc_for_evaluating_of_ev_of_observarble} and \cref{eq:X_measure_for_exp_of_obs}, the above expression could be rewritten as
\begin{align}
    \ev{A}_\text{est} = \frac{\sum_\alpha c_\alpha \pqty{\sum_{i=0}^{n'} g_i^2 \ev{X\otimes I^{\otimes 2}}_{P_\alpha} + 2\sum_{i<j} g_ig_j\ev{X\otimes I^{\otimes 2}}_{P_\alpha}}}{\sum_{i,j=0}^{n'} g_i g_j \ev{X\otimes I^{\otimes 2}}_{I}}.\label{eq:estimator_of_Trotter_se}
\end{align}
Note that the first term of numerator $\ev{X\otimes I^{\otimes 2}}_{P_\alpha}$ depends on $\rho_i$, and the second term of numerator $\ev{X\otimes I^{\otimes 2}}_{P_\alpha}$ and the denominator $\ev{X\otimes I^{\otimes 2}}_{I}$ depend on $\rho_i$ and $\rho_j$.

It is known that a variance of a ratio of two independent random variables $F/G$ could be approximated as
\begin{align}
    \text{Var}\bqty{\frac{F}{G}} = \frac{1}{\ev{G}^2}\bqty{\text{Var}\bqty{F} + \ev{\frac{F}{G}}^2\text{Var}\bqty{G}}.
\end{align}
Using this approximation and \cref{eq:estimator_of_Trotter_se}, we obtain the variance of the estimator:
\begin{align}
    \text{Var}\bqty{A_\text{est}} &= \frac{1}{\pqty{\sum_{i,j=0}^{n'}g_ig_j\ev{X\otimes I^{\otimes 2}}_I}^2}\\
    &\times
    \bqty{\sum_\alpha\sum_{i,j=0}^{n'}c_\alpha^2g_i^4 \text{Var}\bqty{\pqty{X\otimes I^{\otimes 2}}_{P_\alpha}} +4\sum_\alpha\sum_{i<j}c_\alpha^2g_i^2g_j^2 \text{Var}\bqty{\pqty{X\otimes I^{\otimes 2}}_{P_\alpha}}+ \ev{A}_\text{est}^2\sum_{i,j=0}^{n'}g_i^2g_j^2\text{Var}\bqty{\pqty{X\otimes I^{\otimes 2}}_I}}.\label{eq:variance_est_wip}
\end{align}
Substituting \cref{eq:single_shot_variance_of_pauli,eq:single_shot_variance_of_fidelity} into \cref{eq:variance_est_wip}, we obtain
\begin{align}
    \text{Var}\bqty{A_\text{est}} &= \frac{1}{\pqty{\sum_{i,j=0}^{n'}g_ig_j\text{Tr}\pqty{\rho_i \rho_j}}^2}\\
    &\times
    \bqty{\sum_\alpha\sum_{i,j=0}^{n'}c_\alpha^2g_i^4 \pqty{1-\text{Tr}\pqty{\rho_i^2P_\alpha}^2} +4\sum_\alpha\sum_{i<j}c_\alpha^2g_i^2g_j^2 \Bqty{1-\text{Tr}\pqty{\frac{\rho_i\rho_j + \rho_j\rho_i}{2}P_\alpha}^2} + \ev{A}_\text{est}^2\sum_{i,j=0}^{n'}g_i^2g_j^2\pqty{1-\text{Tr}\pqty{\rho_i\rho_j}^2}}.
\end{align}

\section{Shot noise on observable estimation by virtual distillation}\label{sec:single_shot_variance_of_VD}
Here, we give the theoretical analysis of the single-shot variance of the estimator using virtual distillation (VD).
Suppose that we would like to obtain the expectation value of an observable $A = \sum_{\alpha=1}^L c_\alpha P_\alpha$, where $P_\alpha$ is a Pauli operator.
For simplicity, we use two copies of noisy states for VD.
The error-mitigated expectation value with VD is given by
\begin{align}
    \ev{A}_\text{est}
    &= \frac{\text{Tr}\pqty{\rho^2 A}}{\text{Tr}\pqty{\rho^2}},\\
    &=\frac{\sum_{\alpha=1}^{L} c_\alpha \text{Tr}\pqty{\rho^2 P_\alpha}}{\text{Tr}\pqty{\rho^2}}.
\end{align}
Using the quantum circuits shown in \cref{fig:qc_for_evaluating_of_ev_of_observarble} and \cref{eq:X_measure_for_exp_of_obs,eq:X_measure_for_fidelity}, where we set $\rho_i = \rho_j = \rho$, the above expression could be rewritten as
\begin{align}
    \ev{A}_\text{est}
    = \frac{\sum_{\alpha=1}^L c_\alpha \ev{X\otimes I^{\otimes 2}}_{P_\alpha}}{\ev{X\otimes I^{\otimes 2}}_I}.
\end{align}
Similar to \cref{sec:single_shot_variance_of_tse}, the variance of the estimator can be approximated by
\begin{align}
    \text{Var}\bqty{A_\text{est}}
    &=\frac{1}{\ev{X\otimes I^{\otimes 2}}_I^2}\bqty{\sum_{\alpha=1}^Lc_\alpha^2 \text{Var}\bqty{\pqty{X\otimes I^{\otimes 2}}_{P_\alpha}} + \ev{A}_\text{est}^2\text{Var}\bqty{\pqty{X\otimes I^{\otimes 2}}_I}}\\
    &=\frac{1}{\text{Tr}\pqty{\rho^2}^2}\bqty{\sum_{\alpha=1}^Lc_\alpha^2 \pqty{1-\text{Tr}\pqty{\rho^2P_\alpha}^2} + \ev{A}_\text{est}^2\pqty{1-\text{Tr}\pqty{\rho^2}^2}}.
\end{align}

\end{document}